\documentclass[a4paper,11pt]{article}
\pdfoutput=1 
\usepackage[english]{babel}
\usepackage{microtype,xspace}
\usepackage[T2A]{fontenc}
\usepackage[utf8]{inputenc}	

\usepackage{jheppub}
\usepackage{bm,amssymb,amsmath,amsfonts,slashed,graphicx,multirow,soul,mathtools,xspace,array,comment,color}
\hypersetup{colorlinks=true, allcolors=blue}
\usepackage{adjustbox}
\usepackage{siunitx}
\usepackage{bm}
\usepackage{float}
\allowdisplaybreaks
\usepackage{ bbold }
\usepackage{subfigure}
\usepackage[normalem]{ulem}
\usepackage{enumerate}
\usepackage{multirow}
\usepackage{booktabs}
\usepackage{xparse}
\usepackage{etoolbox}
\usepackage{physics}
\usepackage{hypcap}
\usepackage{siunitx}
\usepackage{xcolor}
\usepackage{comment}
\usepackage{gensymb}
\usepackage[capitalise]{cleveref}
\usepackage{chngcntr}

\newcommand{\ldoublet}{l}
\newcommand{\qdoublet}{q}
\NewDocumentCommand{\Op}{ m m O{} o }{
	\O^{\ifblank{#3}{}{#3,}#2 }_{\IfNoValueTF{#4}{#1}{\substack{#1\\#4}}}
}
\NewDocumentCommand{\lwc}{ m m O{} o }{
	L^{\ifblank{#3}{}{#3,}#2 }_{\IfNoValueTF{#4}{#1}{\substack{#1\\#4}}}
}
\NewDocumentCommand{\dlwc}{ m m O{} o }{
	{\dot L}^{\ifblank{#3}{}{#3,}#2 }_{\IfNoValueTF{#4}{#1}{\substack{#1\\#4}}}
}
\NewDocumentCommand{\tlwc}{ m m O{} o }{\widetilde
	L^{\ifblank{#3}{}{#3,}#2 }_{\IfNoValueTF{#4}{#1}{\substack{#1\\#4}}}
}
\NewDocumentCommand{\cwc}{ m m O{} o }{
	C^{\ifblank{#3}{}{#3}#2 }_{\IfNoValueTF{#4}{#1}{\substack{#1\\#4}}}
}

\NewDocumentCommand{\tcwc}{ m m O{} o }{
	\widetilde C^{\ifblank{#3}{}{#3}#2 }_{\IfNoValueTF{#4}{#1}{\substack{#1\\#4}}}
}

\newcommand{\Br}{\text{Br}}
\def\be{\begin{equation}}
\def\ee{\end{equation}}
\newcommand{\TeV}{\text{TeV}}

\newcommand{\fb}{\text{fb}}

\arxivnumber{}

\title{\boldmath  Flavor physics at high-energy muon colliders}

\author[1]{Alfredo Glioti,}
\author[2]{David Marzocca,}
\author[3,4]{Andrea Wulzer}

\affiliation[1]{INFN Sezione di Roma, Piazzale Aldo Moro 2, I-00185 Rome, Italy}
\affiliation[2]{INFN, Sezione di Trieste, Via Bonomea 265, 34136, Trieste, Italy}
\affiliation[3]{Institut de F\'{\i}sica d'Altes Energies (IFAE), The Barcelona Institute of Science and Technology (BIST),
Campus UAB, 08193 Bellaterra, Barcelona, Spain}
\affiliation[4]{ICREA, Instituci\'o Catalana de Recerca i Estudis Avan\c{c}ats, 
Passeig de Llu\'{\i}s Companys 23, 
08010 Barcelona, Spain}

\abstract{\\
Flavor-breaking interactions due to heavy new physics can be probed at a 10\,TeV muon collider in the high-energy production of quarks and leptons. The high collision energy mitigates the suppression of the new interactions, offering sensitivity to interaction scales well above 100\,TeV both in the lepton and in the quark sector. We investigate all possible deformations of the Standard Model that produce quadratic growth  with energy of the four-fermion scattering amplitudes at the muon collider, and we derive sensitivity projections.
Electroweak radiation emission gives access to new observable final states like for instance the production of a charged pair of fermions by the emission of at least one $W$ boson. This, combined with jet flavor tagging, improves the sensitivity and the ability to disentangle different interactions.\\[2pt]
Currently, the best probes of flavor-breaking new physics are high-intensity low-energy measurements of lepton or hadron decays or oscillations. 
The high-energy probes at the muon collider, of the interactions containing a muon bilinear, are competitive with current bounds and strongly superior for some class of transitions.
The sensitivity extends to operators involving the top quark that cannot be tested at low energy.
In addition, muon collider probes are generically less exposed to experimental or theoretical mismodeling uncertainties as they do not target the observation of extremely rare phenomena, nor they rely on extremely accurate measurements and theoretical predictions: the energy enhancement makes the putative effects relatively easy to observe.\\[2pt] 
Muon collider measurements offer a novel pathway towards the exploration of flavor physics at the energy rather than at the intensity frontier.
}

\begin{document}
\maketitle
\flushbottom



\vspace{-6pt}
\section{Introduction}
\label{sec:intro}

The gauge sector of the Standard Model (SM) features a $SU(3)^5$ global symmetry between the three families of quarks and leptons, broken by the Yukawa interactions. Flavor physics is the study of the breaking of this symmetry due to new physics Beyond the SM (BSM). With traditional methods, BSM flavor symmetry breaking is most effectively probed by studying effects that are suppressed or absent in the SM due to its peculiar structure of symmetry breakdown. For instance, the absence of tree-level Flavor-Changing Neutral Currents (FCNC) in the SM makes the study of these transitions an effective probe of BSM sectors with a different pattern of flavor breaking. Similarly, Lepton Flavor-Violating (LFV) processes are forbidden in the SM---up to tiny effects due to neutrino masses---by the accidental conservation of individual lepton numbers, making lepton-flavor changing transitions a powerful BSM probe. Beyond flavor-changing transitions such as FCNC and LFV, flavor-preserving processes are also of interest. One example is the study of Lepton Flavor Universality Violation (LFUV), which is also highly suppressed in the SM. 

The distinction between flavor-changing and flavor-preserving processes is less relevant for the muon collider (MuC) phenomenology. We will readily see that comparable sensitivities are expected to interactions triggering flavor transitions that are rare in the SM and to interactions that produce instead flavor-dependent corrections to abundant SM processes. However, the distinction is still a useful organizing principle especially for comparing the MuC sensitivity with the one of traditional low-energy probes. 

The ongoing design study~\cite{Delahaye:2019omf,AlAli:2021let,Black:2022cth,Accettura:2023ked,InternationalMuonCollider:2024jyv,InternationalMuonCollider:2025sys} could eventually lead to a muon collider with 10\,TeV energy in the center of mass. The relevant processes for our analysis are the direct $\mu^+\mu^-$ annihilation producing a fermion pair with an invariant mass that is nearly as high as the collider energy. The processes of interest are $\mu^- \mu^+ \to f \bar{f}^{(\prime)}$, with $f\bar{f}^{(\prime)}$ two quarks or leptons of either same or different flavor. Notice that $f$ and $f^{\prime}$ do not have necessarily the same electric charge because the charge mismatch with the initial state can be compensated by the emission of charged unresolved $W$ bosons. Previous studies of high-energy difermion (and diboson or boson-Higgs) production at the muon collider, targeting both flavor-preserving and flavor-breaking heavy new physics, include~\cite{Buttazzo:2020uzc,Chen:2022msz,Huang:2021biu,Azatov:2022itm,Altmannshofer:2023uci,Ake:2023xcz,Homiller:2022iax,Li:2023lin,Sun:2023cuf,Bhattacharya:2023beo,Chang:2024xvc} and are reviewed in~\cite{Accettura:2023ked,InternationalMuonCollider:2025sys}.

For a first assessment of the muon collider difermion measurements potential, consider a neutral flavor-changing transition such as a FCNC in the quark sector or a LFV process, mediated by a dimension-six interaction with Wilson coefficient $1/{\Lambda_{\rm FC}^2}$. The new physics contribution to the scattering amplitude is of order $E_{\rm{cm}}^2/{\Lambda_{\rm FC}^2}$, where $E_{\rm{cm}}$ is the collider energy. Neutral flavor-changing transitions are extremely rare in the SM. Therefore the background emerges from a flavor-preserving SM process that fakes a flavor-changing final state due to, for instance, the mis-identification of one of the flavors. By denoting as $\epsilon_{\rm f}$ the relevant fake rate and estimating as $e^2\sim1/10$ (with $e$ the electric charge) the SM amplitude for the flavor-preserving process due to Electroweak (EW) interactions, the contribution to the cross section from the flavor-changing new physics, relative to the SM background, reads
\be
    \frac{\delta \sigma_{\rm FC}}{\sigma_{\rm bkg}} \approx \frac{1\%}{\epsilon_{\rm f}} \left(\frac{E_{\rm cm}}{10~\TeV}\right)^4 \left(\frac{100~\TeV}{\Lambda_{\rm FC}}\right)^4.
    \label{eq:estimate_FV_EFT0}
\ee
For $\Lambda_{\rm FC}=100$\,TeV, a large order $1\%$ effect is expected at the 10\,TeV MuC even with unrealistically poor flavor tagging performances $\epsilon_{\rm f}=1$. 

An accuracy well below $1\%$ in the experimental measurement of the cross section and in its theoretical prediction can be safely assumed, therefore the muon collider sensitivity to new interactions with about 100\,TeV scale is dictated by statistics. The typical flavor-preserving cross section is around $1$~fb at the 10\,TeV MuC. With the target luminosity of $10~{\textrm{ab}}^{-1}$, this corresponds to $10^4\cdot\epsilon_{\rm f}$ mis-identified background events, i.e.,\ to $1\%/\sqrt{\epsilon_{\rm f}}$ statistical precision. The estimate is independent of the MuC energy because the $1/E_{\rm cm}^2$ scaling of the cross section is compensated by the $E_{\rm cm}^2$ scaling~\cite{Delahaye:2019omf,AlAli:2021let,Black:2022cth,Accettura:2023ked,InternationalMuonCollider:2024jyv,InternationalMuonCollider:2025sys} of the collider luminosity. Comparing with Eq.~(\ref{eq:estimate_FV_EFT0}) gives the $1\sigma$ sensitivity estimate 
\be
\Lambda_{\rm FC}^{(1\sigma)}\approx\frac{100~\TeV}{(\epsilon_{\rm f})^{1/8}}\frac{E_{\rm cm}}{10~\TeV}\,.
    \label{eq:estimate_FV_SENS}
\ee
The 10\,TeV muon collider is sensitive to the 100\,TeV interaction scale, and the reach increases linearly with the collider energy. Flavor tagging can improve the sensitivity above 100\,TeV, taking however into account the weak dependence of the reach on the eighth root of the fake probability $\epsilon_{\rm f}$.

Interestingly enough, the muon collider features comparable sensitivity to interactions that do not mediate a change of flavor. Consider for instance a  LFUV interaction, or a violation of universality in the quark sector. If the flavor of the fermion involved in the interaction (e.g., a lepton or a heavy quark) can be tagged, the muon collider can test the new interaction in the $f \bar{f}$ final state by comparing with the flavor-universal SM predictions or with final states $f^\prime \bar{f}^\prime$ of a different flavor $f^\prime$. The production of same-flavor $f \bar{f}$ is abundant in the SM, with a typical cross section that corresponds to $1\%$ statistical precision as previously discussed. The universality-breaking (but flavor-preserving) new interaction gives a contribution of order $E_{\rm{cm}}^2/{\Lambda_{\rm FP}^2}$ to the scattering amplitude, and $e^2\sim1/10$ is a good estimate of the SM amplitude. Depending on whether the helicity structure of the new interaction interferes with the SM or not, we obtain---see later Eq.~(\ref{eq:sigma_EFTeffect})---sensitivity estimates
\be
\Lambda_{\rm FP}^{(1\sigma)}\overset{\rm{w/~int.}}{\approx}{320~\TeV}\frac{E_{\rm cm}}{10~\TeV}\,,\qquad\;
\Lambda_{\rm FP}^{(1\sigma)}\overset{\rm{w/o\,int.}}{\approx}{100~\TeV}\frac{E_{\rm cm}}{10~\TeV}\,,
    \label{eq:estimate_FP_SENS}
\ee
that are again in the 100\,TeV ballpark. This motivates the comprehensive investigation of high-energy difermion production processes that is the subject of the present paper.

The emission of massive EW bosons in addition to the hard fermion pair is enhanced by large Sudakov logarithms at the 10\,TeV MuC and this further enlarges the set of relevant observables~\cite{Chen:2022msz}. For instance, the emission of a soft unresolved charged $W$ boson produces a hard $f \bar{f}^{\prime}$ final state with non-vanishing total charge, at a rate that is only slightly smaller than the one of the corresponding neutral process. The hard four-fermion scattering amplitude for the production of the charged final state is sensitive to different BSM interactions than the neutral amplitude. Charged cross section measurements offer sensitivity to new interactions enabling a more complete exploration of new physics. EW radiation can be exploited also for neutral $f \bar{f}$ production by combining cross section measurements in `exclusive' and in `with-radiation' final states where EW radiation emission is, respectively, vetoed or required. New physics contributes differently to these observables, enhancing the sensitivity and the ability to disentangle different interactions. In particular, the combined measurements offer new handles to probe the chirality structure of the new interactions because EW radiation emission is sensitive to chirality. 

Considering exclusive and with-radiation neutral processes, charged processes and flavor tagging both in the quark and in the lepton sector results in a large number of cross sections and angular distributions that can be measured at a muon collider. In this paper we investigate the perspectives of these measurements to probe the most general Lorentz-invariant deformation of the SM amplitudes that scales quadratically with the energy. These deformations are conveniently parametrized as dimension-six four-fermion operators in the SM Effective Field Theory (EFT) involving two second-generation leptons (muon or muon-neutrino). The sensitivity projections on the EFT interaction scale is compared with current knowledge and future prospects of traditional low-energy measurements of lepton or hadron decays.

The rest of the paper is organized as follows. In Section~\ref{sec:dif} we describe the final states and the measurements employed in the analysis. The theoretical predictions are based on~\cite{Chen:2022msz} and include the resummation of the leading Sudakov EW double logarithms. In the diquark final state, the observable cross sections are obtained by a simple strategy based on parametrized jet flavor tagging and mistagging rates, which we estimate based on CLIC detector studies~\cite{CLIC:2018fvx} and current LHC experiments performances. The background for some of the dilepton final states needs instead to be estimated by simulations.

Section~\ref{sec:FC} reports our results for operators mediating FCNC in the quark sector, LFV and LFUV in the lepton sector. The sensitivity is compared with current knowledge and, when available, with future prospects of traditional flavor physics measurements. Several examples are provided of different muon collider measurements (in, e.g., the neutral and the charged final state) probing complementarity directions in the space of Wilson coefficients. 
In Section~\ref{sec:conclusions} we report our conclusions and an outlook to future work. Appendix~\ref{app:EWresum} describes the basic formulas for the Sudakov double logs from~\cite{Chen:2022msz}, whose explicit implementation is publicly available~\cite{Glioti:2025}. Appendix~\ref{app:lowEflavor} summarizes the low-energy flavor physics observables and the corresponding constraints on some of the flavor transitions studied in the paper. Appendix~\ref{app:tab} collects some of the tables described in the main text.

\section{High-energy difermion production at muon colliders}
\label{sec:dif}

Colliding elementary particles such as the muons allows to take advantage of the full collider energy in the hard scattering process. In this work we focus on the production of two fermions of either same or different flavor and charge from direct muon anti-muon annihilation:
\be\label{eq:procs}
\mu^- \mu^+ \to f \bar{f}^{(\prime)}+X\,.
\ee
The relevant kinematic regime is when the final state fermions are nearly as energetic as the muon beam, consider for definiteness $E(f),\,E( \bar{f}^{(\prime)}) > 0.4\cdot  E_{\textrm{cm}}$\,, central and back-to-back. In Eq.~(\ref{eq:procs}), $X$ denotes a generic multi-particle state which is relatively soft since its total energy that must be less than $0.2\cdot  E_{\textrm{cm}}$ by energy conservation. At the leading double logarithm order (see below) these particles can only be photons, gluons or EW bosons.

The tree-level predictions for the difermion production process~(\ref{eq:procs}) are subject to virtual and real EW corrections that are enhanced by large Sudakov Double Logarithms (DL). The DL corrections scale as $g^2/16\pi^2\,\log^2{E_{\textrm{cm}}^2/m_{\textrm{w}}^2}$ at one loop---with $g$ the SU$(2)_L$ coupling and $m_{\textrm{w}}$ the $W$ boson mass---times a Casimir factor that can be as large as 3 (depending on the external states polarization) in di-fermion production. Compatibly with this estimate, the EW DL are of order one for $E_{\textrm{cm}}=10$\,TeV and cannot be treated at fixed order in perturbation theory. We base our predictions---described in Appendix~\ref{app:EWresum}---on the Sudakov DL resummation of~\cite{Chen:2022msz}. We consider two kinds of observables for which resummation is possible: the \emph{exclusive} and the \emph{semi-inclusive} cross section. The two classes of cross section are characterized by different requirements on the real radiation $X$ that accompanies---see Eq.~(\ref{eq:procs})---the emission of the hard fermion pair. 

An \emph{exclusive} final state is one where a veto is applied excluding the presence in the final state of massive EW bosons or hard photons.\footnote{We veto photons with a hardness (defined, e.g./ as the minimal transverse momentum relative to the hard particles) above the EW scale $m_{\textrm{w}}\simeq 80$~GeV.} Namely, the radiation $X$ in Eq.~(\ref{eq:procs}) consists only of soft or collinear photons and gluons. In the exclusive case, the DL resummation can be performed at the amplitude level and is known since long~\cite{Fadin:1999bq}. It amounts to a rescaling of the tree-level cross section by a multiplicative factor that depends on the EW quantum numbers of the external states as in Eq.~\eqref{eq:exclSudakov}. The DL resummation reduces the tree level cross section by a factor that can reach even $50\%$ at the 10\,TeV MuC for some processes.

A \emph{semi-inclusive} final state is instead a one where no veto is applied and an arbitrary pattern of radiation is allowed. The final state is only partly (i.e., semi-) inclusive because the EW color of the final state fermions is selected, while it would be summed over in a `fully-inclusive' process. The resummation of the semi-inclusive cross section cannot be performed at the amplitude level because the real radiation emission mixes up amplitudes with different EW color on the external legs. The resummed expressions~\cite{Chen:2022msz} are obtained for the density matrix in the space of EW color, as in Eq.~(\ref{eq:resSI0}). 

The difference between the semi-inclusive and the exclusive cross section for the same hard fermion pair is called the \emph{with-radiation} cross section. It corresponds to a final state where the presence of massive EW bosons or hard photons is required, compatibly with the upper bound on the total energy of the radiation particles $X$. The with-radiation final state has no overlap with the exclusive final state. The corresponding cross section measurements are the statistically independent observables that we employ in our fit. Interesting with-radiation processes are the production of a hard fermion pair with non-vanishing total charge, i.e., different charges for $f$ and $f^\prime$ in Eq.~(\ref{eq:procs}). Such processes only emerge in with-radiation final states because they require the emission of at least one charged $W$ boson. The large Sudakov DL enhances the cross section of such charged processes, which is typically comparable with the one of the neutral processes at the 10\,TeV MuC. 

We study three classes of processes in each possible flavor configuration: exclusive and with-radiation neutral processes, plus charged (with-radiation) processes. The processes are described in the two following sections for diquark and dilepton production, respectively. The resummed predictions employed in the analysis are implemented in a public code~\cite{Glioti:2025}, including the effect of the new EFT interactions listed in Section~\ref{sec:FC}.

\subsection{Diquark production}

Using jet flavor tagging, the production of any hard pair of quarks---i.e., the process~(\ref{eq:procs}) with $f,f^\prime=q,q^\prime$---contributes to one of the following categories:
\begin{equation}
    2b, \quad 2c, \quad 2j, \quad b+j, \quad c+j, \quad b+c, 
    \quad 2 t^{\cal B}, \quad  t^{\cal B} + b, \quad  t^{\cal B} + c, \quad t^{\cal B} + j~,
    \label{eq:tag_categories}
\end{equation}
where $j$ is a light jet (i.e., a jet that is not tagged as $b$ or $c$) and $t^{\cal B}$ is a boosted top quark. For each flavor-tagged category, the observables are the exclusive and the with-radiation cross section (and angular distribution) as previously explained, with the exception of the $b+c$ and $t^{\cal B} + b$ categories where only the with-radiation cross section is considered. This is because the $b+c$ and $t^{\cal B} + b$ final states only emerge---up to mistagging---from charged diquark production that requires the emission of a $W$ boson. Notice that we are ignoring the possibility to tag the strange quark and the one of discriminating the charge of the quarks. It would be interesting to explore the impact of strange and charge tagging (if feasible) on our results, but this is left to future work.

The relevant event topology is the one where the two jets are very energetic and back-to-back, which could be enforced for instance by a $E_{\textrm{min}}=0.4\cdot  E_{\textrm{cm}}=4\,$TeV selection cut on the jets energy, plus a large angular separation requirement. The exact cuts to be employed in the analysis are not relevant in our simplified treatment because the soft-collinear radiation emission does not affect the Born-level kinematic considerably. Therefore, our predictions are insensitive to the cuts at the DL order and single logarithm effects are under control provided the $E_{\textrm{min}}$ cut is not too stringent~\cite{Chen:2022msz}. Large single logarithms emerge if the jets are emitted in the forward/backward direction. For this reason, we limit our analysis to the angular region $\theta\in[20^\circ,160^\circ]$ instead of exploiting the entire coverage of the MuC detector.

We parametrize jet tagging efficiencies and mistag rates by the probabilities of true quarks being tagged to have a certain flavor, which form an efficiency/mistag matrix $\epsilon_{\rm true}^{\rm tagged}$. Employing values inspired by CLIC detector studies~\cite{CLIC:2018fvx}, we set
\begin{equation}\begin{split}
    & \epsilon_b^b = 0.8, \quad
    \epsilon_c^b = 0.1, \quad
    \epsilon_{s,d,u}^b = 0.01, \\
    & \epsilon_b^c = 0.1, \quad
    \epsilon_c^c = 0.5, \quad
    \epsilon_{s,d,u}^c = 0.02.
    \label{eq:jet_tag_eff}
\end{split}
\end{equation}
These numbers are very conservative in comparison with the state-of-the-art flavor tagging in LHC experiments \cite{Bols:2020bkb,ATLAS:2025dkv} and with FCC-ee projections~\cite{Bedeschi:2022rnj,Blekman:2024wyf}. However, LHC and FCC-ee performances are not directly portable to the multi-TeV energy regime that is relevant for our analysis. MuC detector studies are needed for a robust estimate of the flavor tagging performance parameters.
Nevertheless, notice that our results are very weakly sensitive to the specific values we employed because the sensitivity reach on the EFT scale $\Lambda$ scales with the eight root power of the tagging efficiencies and mistag rates, as shown in Eq.~\eqref{eq:estimate_FV_SENS}.

The entries of the efficiency/mistag matrix that involve the top quark are estimated based on the ATLAS analysis of Ref.~\cite{ATLAS:2018wis}, which considers boosted tops with $p_T \in [1.5, \,2]\,\TeV$ (see also \cite{ATLAS:2024rua}). For the top tagging efficiency and the rejection rates from the light quarks (including the $b$), we consider the following working point:
\begin{equation}
    \epsilon_t^{t^{\cal B}} = 0.6, \quad
    \epsilon_{b,c,s,d,u}^{t^{\cal B}} = 0.04~.
\end{equation}
We could not find in the literature an estimate of the probability of a true top quark to be tagged as a $b$, $c$ or light jet.
We set them by assuming that the top quarks that do not pass the boosted top selections have an $\epsilon_b^b$ probability to be tagged as $b$-jets, that is $\epsilon_t^b = (1-\epsilon_t^t)\epsilon_b^b$. The analogous probabilities of a top to be mistagged as $c$ or light jets are small and we set them to zero. In summary:
\begin{equation}
    \epsilon_t^b = 0.3, \quad
    \epsilon_t^{c,j} = 0~.
\end{equation}
Note that the probabilities of a true top quark be reconstructed as a jet of any flavor do not sum up to one. The mismatch can be attributed to the probability of the top being reconstructed as a hadronic vector boson or Higgs.
Finally, we fix the light jets efficiencies/mistags by imposing that the probabilities sum to one, $\epsilon_{q}^j = 1 - \epsilon_q^b - \epsilon_q^c - \epsilon_q^{t^{\cal B}}$:
\begin{equation}
    \epsilon_b^j = 0.06, \quad
    \epsilon_c^j = 0.36, \quad
    \epsilon_{s,d,u}^j = 0.93.
\end{equation}

\begin{table}[t]
\centering
\setlength\extrarowheight{-2pt}
\begin{tabular}{|c|cc|cc|}
\hline
& \multicolumn{2}{c|}{3\,TeV} & \multicolumn{2}{c|}{10\,TeV} \\
\hline
$\sigma_\text{SM}$ [fb]& Excl & Rad & Excl & Rad \\
\hline
$2j$ & 24 & 10 & 1.7 & 1.4 \\
$c+j$ & 5.8 & 3.5 & 0.42 & 0.47 \\
$b+j$ & 2.0 & 1.0 & 0.14 & 0.14 \\
$t^{\cal B}+j$ & 2.3 & 1.2 & 0.16 & 0.16 \\
$2c$ & 3.4 & 0.64 & 0.25 & 0.093 \\
$b+c$ & --- & 0.61 & --- & 0.087 \\
$t^{\cal B}+c$ & 0.62 & 0.51 & 0.045 & 0.067 \\
$2b$ & 5.2 & 2.2 & 0.37 & 0.30 \\
$t^{\cal B}+b$ & --- & 3.4 & --- & 0.45 \\
$2t^{\cal B}$ & 4.8 & 0.98 & 0.35 & 0.14 \\
\hline
\end{tabular}\qquad
\caption{Exclusive and with radiation SM cross sections for diquark final states, for 3 and 10\,TeV muon colliders, integrated in the angular region between $20^\circ$ and $160^\circ$.
}
\label{tab:xsecsquarks}
\end{table}

Using the tagging efficiencies we obtain the predicted differential cross sections for each of the categories listed in Eq.~\eqref{eq:tag_categories}:
\begin{eqnarray}
    \frac{d \Sigma({\mathfrak{f}}_A {\mathfrak{f}}_B)}{d|\!\cos\theta|} = k_{AB}  \sum_{i,j}&&\left[ 
    (\epsilon^{{\mathfrak{f}}_A}_{d_i} \epsilon^{{\mathfrak{f}}_B}_{d_j} + \epsilon^{{\mathfrak{f}}_B}_{d_i} \epsilon^{{\mathfrak{f}}_A}_{d_j}) 
       \, \frac{d \sigma(d_i \bar{d}_j)}{d |\!\cos\theta|} +
    (\epsilon^{{\mathfrak{f}}_A}_{u_i} \epsilon^{{\mathfrak{f}}_B}_{u_j} + \epsilon^{{\mathfrak{f}}_B}_{u_i} \epsilon^{{\mathfrak{f}}_A}_{u_j}) 
       \, \frac{d \sigma(u_i \bar{u}_j)}{d |\!\cos\theta|} + \right.\nonumber\\
     &&\left. 
     +
    (\epsilon^{{\mathfrak{f}}_A}_{d_i} \epsilon^{{\mathfrak{f}}_B}_{u_j} + \epsilon^{{\mathfrak{f}}_B}_{u_j} \epsilon^{{\mathfrak{f}}_A}_{d_i}) 
       \, \left( \frac{d \sigma(d_i \bar{u}_j)}{d |\!\cos\theta|} + \frac{d \sigma(u_j \bar{d}_i)}{d |\!\cos\theta|} \right) \right].
\label{eq:x_seg_tag}
\end{eqnarray}
In the equation, ${\mathfrak{f}}_{A,B}=t^{\cal B},b, c, j$ denotes the tagged flavor and $k_{AB}=1/(1+\delta_{AB})$. The sum runs over the quark generation indices $i,j=1,2,3$ and $u_i$ ($d_i$) are the up- (down-) type quarks of each generation. The cross sections (either exclusive or with-radiation) for di-quark productions are obtained by including the Sudakov logarithms in the tree-level cross sections through the analytic formulas in Appendix~\ref{app:EWresum}. Notice that the resummed formulas depend on the hard Born-level amplitudes, which are readily computed either in the SM or including the contribution of the new EFT operators listed in Section~\ref{sec:FC}.

In our analysis we exploit angular information to improve the sensitivity and the power to discriminate different operators. Since the jets are nearly back-to back, and the quark charge is not tagged, the only available angular variable is the angle between the jet and the beam axes, which ranges from 0 to 90$^\circ$. The cosine of this angle is denoted as $|\!\cos\theta|$ in Eq.~(\ref{eq:x_seg_tag}). Five equally-spaced bins in $|\!\cos\theta|$ are considered, ranging from 0 to $\cos20^\circ\simeq0.94$. 

The SM predictions for all the exclusive and with-radiation diquark cross sections integrated in the central region are reported in Table~\ref{tab:xsecsquarks}. At the 10\,TeV MuC, the exclusive and with-radiation cross section in each flavor-tagged category are comparable, due to the large Sudakov double logarithms that boost real EW radiation emission. With the 10~ab$^{-1}$ target luminosity of the 10\,TeV MuC, the yield for the unsuppressed (flavor-preserving) categories is about $10^4$ events. This corresponds to percent-level measurement accuracy and entails order 100\,TeV scale sensitivity for flavor-diagonal EFT interactions.

While the 10\,TeV MuC is our main target, we also study the reach of a possible first MuC stage with $E_{\rm{cm}}=3$\,TeV and 1~ab$^{-1}$ integrated luminosity. The cross sections---reported in Table~\ref{tab:xsecsquarks}---scale approximately as $1/E_{\rm{cm}}^2$ and are about 10 times larger than at 10\,TeV. The event yields and in turn the precision of the measurements are similar to the ones of the 10\,TeV MuC because of the reduced luminosity. The EFT scale sensitivity thus lowers in proportion to the reduced collider energy and it is estimated at about 30\,TeV. Notice that at 3\,TeV there is a larger gap between the exclusive and the with-radiation cross section because the Sudakov double logarithm is smaller. 

We stress that ours is a rather crude treatment of flavor tagging. Apart from the aforementioned large uncertainty on the entries of the efficiency/mistag matrix, the usage of Eq.~(\ref{eq:x_seg_tag}) only accounts for the diquark processes contribution to the flavor-tagged dijet categories, ignoring other sources of background. For instance, an energetic massive vector boson can fake a boosted top, or even a light jet if it decays asymmetrically producing one quark that is very energetic while the other one is soft. In spite of these limitations, our simplified approach is justified because flavor tagging performances are not the key enabling factor for our analysis and the sensitivity to the EFT scale is weakly sensitive to the fake rate and to the addition of background. It should be  noted that the energetic dijet final state that we target can be mimicked only by hard processes, and the cross section for any such process---such as for instance diboson production from direct $\mu^+\mu^-$ annihilation---does not exceed about 1~fb ($0.1$~fb) at the 10\,TeV (3\,TeV) MuC. Therefore, since the probability of a dijet fake is small, the neglected backgrounds are surely not overwhelming in comparison with the processes we included. Nevertheless, a fully quantitative assessment of the sensitivity requires detailed detector simulations that should soon become available following the ongoing advances of the MuC detector design study~\cite{InternationalMuonCollider:2025sys,MAIA:2025hzm}.

\subsection{Dilepton production}
\label{sec:dilepton}

In case of leptons, we consider all flavor combinations of two opposite-charge leptons with a tagged flavor $\ell=e,\mu,\tau$, plus the charged processes producing a lepton and a neutrino: 
\begin{equation}
\label{dilproc}
\begin{split}
    &\hspace{4pt}{\textrm{Neutral $\ell\ell$:}}\qquad\mu^- \mu^+, \quad
    e^- e^+, \quad
    \tau^- \tau^+, \quad
    e^\mp \mu^\pm, \quad
    \tau^\mp \mu^\pm, \quad
    e^\mp \tau^\pm  \\
    &{\textrm{Charged $\ell\nu$:}}\qquad e^\pm \nu , \quad
    \mu^\pm \nu , \quad
    \tau^\pm \nu.
\end{split}
\end{equation}
Both the exclusive and the with-radiation cross section measurements are considered for the neutral same-flavor processes, while the charged processes can be only of the with-radiation type. Only exclusive measurements are considered for neutral processes with different flavors, for reasons that will become momentarily clear. As for diquark production, the event topology of interest features very energetic back-to-back leptons. A lower cut of $0.4\cdot  E_{\textrm{cm}}$ is applied to the energy of electrons and muons. Additional selections needed in some of the final states---in order to deal with the tau leptons and the missing momentum from the neutrinos---are described later.

\begin{table}[t]
\centering
\begin{tabular}{|c|ccc|ccc|}
\hline
& \multicolumn{3}{c|}{3\,TeV} & \multicolumn{3}{c|}{10\,TeV} \\
\hline
$\sigma_\text{SM}$ [fb]& Excl & Rad & Bkg & Excl & Rad & Bkg \\
\hline
$ee$ & 8.5 & 1.7 & 0 & 0.62 & 0.25 & 0 \\
$\mu\mu$ & 682 & 141 & 0 & 50 & 20 & 0 \\
$\tau\tau$ & 2.3 & 0.47 & 0 & 0.17 & 0.068 & 0 \\
$e^- \bar\nu$\,/\,$\nu e^+$ & --- & 0.96 & 0 & --- & 0.12 & 0 \\
$\mu^- \bar\nu$\,/\,$\nu \mu^+$ & --- & 87 & 0 & --- & 11.3 & 0 \\
$\tau^- \bar\nu$\,/\,$\nu \tau^+$ & --- & 0.48 & 0 & --- & 0.062 & 0 \\
$\mu^-\tau^+$\,/\,$\tau^-\mu^+$ & 0 & --- & 0.21 & 0 & --- & 0.028 \\
$e^-\tau^+$\,/\,$\tau^-e^+$ & 0 & --- & 0.12 & 0 & --- & 0.011 \\
$\mu^- e^+$\,/\,$e^- \mu^+$ & 0 & --- & 0.26 & 0 & --- & 0.029 \\
\hline
\end{tabular}
\caption{SM cross sections of the dilepton final states, for the 3 and the 10\,TeV MuC, integrated in angle between $20^\circ$ and $160^\circ$. For each collider energy, the first two columns report the exclusive and with-radiation SM cross sections. The reducible background, reported in the third column, is considered as a background to the exclusive measurement. The with-radiation measurement is excluded from the analysis in those final states where the reducible background is not negligible.
}
\label{tab:xsecsleptons}
\end{table}

Unlike for jets, the charge of the leptons can be measured enabling the complete determination of the scattering angle $\theta$. We exploit this variable for a binned analysis with 10 equally-spaced bins in $\cos\theta$, restricted to the central region $\theta\in[20^\circ,160^\circ]$ where the single-logarithm corrections are under control. The dilepton cross sections include the Sudakov logarithms as in Appendix~\ref{app:EWresum} and the contribution from the four-leptons operators introduced in Section~\ref{sec:FC}.

The cross section of the flavor-preserving final states and of the charged processes is large in the SM and it dominates over the reducible backgrounds, which are therefore negligible. On the other hand, flavor-changing final states are forbidden in the SM and reducible backgrounds must be taken into account. The main sources are $\tau\tau$ production, which fakes a flavor-changing final states after $\tau$ decays, $WW$ production with leptonic decay and other non-leptonic processes. The non-leptonic processes are a pure background because they are not affected by the EFT operators. New physics contributions are instead present in the $\tau\tau$-induced reducible backgrounds to flavor-changing final states, but they are safely negligible.

The estimate of the cross section is presented below for each final state. The corresponding SM predictions, integrated in the central region, are listed in Table~\ref{tab:xsecsleptons}. In case of the flavor-changing final states we also report,
in Table~\ref{tab:LFVbkg}, the reducible background cross sections in each $\cos \theta$ bin.

\paragraph{$\boldsymbol{ee}$ and $\boldsymbol{\mu\mu}$:} The production of $e^+e^-$ or $\mu^+\mu^-$ is abundant in the SM, and provides the main background for the detection of BSM effects in these final states. Furthermore, we can safely assume unit reconstruction efficiency for electrons and muons. Therefore, the predictions for these final states are merely equal to the corresponding truth-level production cross sections.

\paragraph{$\boldsymbol{e\nu}$ and $\boldsymbol{\mu\nu}$:} This final state can be effectively selected by requiring a large (about $0.4\cdot  E_{\textrm{cm}}$) amount of missing energy, in addition to the energetic lepton. A large angular separation between the lepton and the missing 3-momentum could be also imposed to improve the purity of the signal. The SM production of lepton-neutrino is the dominant background, therefore no reducible background needs to be considered. Notice that since the flavor of the neutrino cannot be tagged, the $e\nu$ final state is potentially sensitive to BSM flavor-changing processes such as $e{\nu}_\mu$ or $e{\nu}_\tau$ production and not only to $e{\nu}_e$ production (and similarly for the $\mu\nu$ final state). However, these final states are not included in our analysis of lepton flavor violating operators. In the absence of these operators, and since the muon and electron efficiencies are large, also for the $e\nu$ and $\mu\nu$ final states our predictions are equal to the corresponding truth-level cross sections.

\paragraph{$\boldsymbol{\mu{e}}$:} The process is forbidden in the SM, therefore the (reducible) background emerges from a more complex final state where some of the particles are lost, the natural candidate being $\mu^-e^+{\nu}_e\bar{\nu}_\mu$ (and similarly for the $\mu^+e^-$ final state). For an estimate, we simulate the process with \texttt{MadGraph5\_aMC@NLO}~\cite{Alwall:2014hca} obtaining (see Table~\ref{tab:xsecsleptons}) a cross section of about $0.029$~fb at the 10\,TeV MuC ($0.26$~fb, at 3\,TeV) with the $0.4\cdot  E_{\textrm{cm}}$ cut on the leptons energy and in the central region $\theta\in[20^\circ,160^\circ]$. The dominant topologies are the direct production of $W^+W^-$ with leptonic decay, and the scattering $\mu^-W^+\to \mu^-W^+$ (with $W^+\to e^+\nu_e$) from an effective $W^+$ emitted by the incoming $\mu^+$. The two topologies account for approximately $1/3$ and $2/3$ of the background cross section, respectively. Instead, the contribution from $\tau^+\tau^-$ production with leptonic tau decays is negligible after cuts.
Based on the simulation, the reducible background component to the cross section is estimated in each analysis bin (see Table~\ref{tab:LFVbkg}), and combined with the new physics contribution from dilepton production.

It should be noted that our tree-level estimate of the reducible background is subject to large Sudakov corrections. Moreover, the exclusive and semi-inclusive cross sections are equal at the tree-level order such that the with-radiation background cross section (which is the difference between the two) is formally zero. For this reason we exclude the with-radiation measurement and we only consider the exclusive cross section determination for the $\mu{e}$ final state. Similar considerations hold for the $\tau{e}$ and $\tau\mu$ final states. Including the Sudakov effects in the reducible background is possible in principle, but it goes beyond the exploratory scope of the present work. 

\paragraph{$\boldsymbol{\tau{e}}$:} We consider hadronic decays of the tau lepton, hence the observed final state is a low-multiplicity jet---which we take within the central angular region---from the visible tau decay products, plus the lepton. We simulated the hadronic decays of the $\tau$ using the \texttt{TauDecay} UFO model~\cite{Hagiwara:2012vz}. We apply a $0.4\cdot  E_{\textrm{cm}}$ lower cut on the electron energy and a looser cut $E_{\textrm{vis}}>0.2\cdot  E_{\textrm{cm}}$ on the visible products of the tau decay. On the `signal' $\tau{e}$ production process, the $E_{\textrm{vis}}$ selection cut has an efficiency of about $80\%$ , which we estimate by a \texttt{MadGraph} simulation where $\mu\mu\to\tau{e}$ is induced by a suitable four-lepton EFT operator. Taking also into account the hadronic $\tau$ branching ratio of $65\%$, this gives a total efficiency of $50\%$ that we apply to the truth-level cross section prediction. 

The dominant reducible background is the $W^+W^-$ process with leptonic $W$ decays producing the electron and the tau. A significant reduction of this background can be obtained by exploiting the following observation. In the signal, the momentum of the $\nu_\tau$ from the $\tau$ decay can be approximately reconstructed because it is the main source of missing four-momentum, $p_{\textrm{miss}}$. Since the $\tau$ is boosted, the $\nu_\tau$ is collinear to the visible products of the $\tau$ decay and therefore a mild separation cut $\Delta{R}(p_{\textrm{miss}},p_{\textrm{vis}})<1$ has unit efficiency on the signal topology.\footnote{$\Delta R$ is defined as $\sqrt{\Delta\eta^2+\Delta\phi^2}$ as customary at hadron colliders.} Many additional sources of missing momentum are present in the background. Therefore $p_{\textrm{miss}}$ and $p_{\textrm{vis}}$ are not collinear and $\Delta{R}<1$ is effective for background reduction. After all the cuts, the $WW$ background has a cross section of about $0.0085$~fb at the 10\,TeV MuC ($0.095$~fb at 3\,TeV). 
A second important background is the production, in the SM, of $\tau^+\tau^-$ with the leptonic decay of one tau. The tree-level cross section after cuts is $0.0024$~fb at the 10\,TeV MuC ($0.027$~fb at 3\,TeV). 

\paragraph{$\boldsymbol{\tau{\mu}}$:} The selection cuts are the same as in the $\tau{e}$ final state, therefore the same estimates apply for the efficiency factor and for the background components from $WW$ and $\tau\tau$ production. In addition, the $\tau\mu$ final state features a new source of reducible background from effective $W$ scattering with the muon: $\mu^\mp W^\pm\to \mu^\mp W^\pm$ with $W^\pm\to \tau^\pm\nu_\tau$. The cross section after cuts of all backgrounds to this process is $0.028$~fb at the 10\,TeV MuC and $0.21$~fb at 3\,TeV.

\paragraph{$\boldsymbol{\tau\nu$}:} This final state can be selected by the $E_{\textrm{vis}}>0.2\cdot  E_{\textrm{cm}}$ cut on the visible products of the hadronic $\tau$, plus the requirement of large missing energy. The cut has a $80\%$ efficiency on the direct $\tau\nu$ production process and including the hadronic tau branching ratio sets an overall efficiency of $50\%$. No contribution is considered from $\tau\nu_e$ and $\tau\nu_\mu$ production because $\tau\nu$ final state measurements are not used to probe lepton flavor violating operators.

\paragraph{$\boldsymbol{\tau\tau$}:} We request the $E_{\textrm{vis}}>0.2\cdot E_{\textrm{cm}}$ cut on the visible energy of each tau jet. This has an efficiency of $65\%$ on the direct $\tau^+\tau^-$ production process. Multiplying by the square of the hadronic tau branching ratio we obtain a total efficiency factor of $27\%$. The dominant background is the SM $\tau^+\tau^-$ and no reducible background needs to be considered.

\section{Sensitivity to flavor-breaking new physics}
\label{sec:FC}

Our goal is to assess the MuC sensitivity to flavor-breaking heavy new physics through the study of the difermion processes described in the previous section.  If the BSM particles are above the collider energy, \( M_{\rm BSM} > E_{\rm cm} \), their effects emerge as deformations of the SM scattering amplitudes and are described systematically within an EFT framework. The deformations of interest are those that scale quadratically with the energy and that involve two second-generation leptons. Following the Warsaw basis classification~\cite{Grzadkowski:2010es}, they correspond to the four-fermion EFT operators listed in Table~\ref{tab:SMEFT_operators}. In the table, $\ldoublet$ and $\qdoublet$ denote the lepton and quark SM doublet left-handed fields, while $e$, $u$ and $d$ are the right-handed singlets. We employ the down-quark mass basis, in which the left-handed quark doublet is given by \( q^i_{\textsc{m}} = (V^*_{ji} u_L^j, d_L^i) \), with \( V \) the CKM matrix. Generation indices are not listed explicitly in the table. They will be reported when needed as upper indices, in the form of ${\mathcal{O}}^{[ijkl]}$. It is understood that the relevant operators contain at least 2 second-generation $\ldoublet$ or $e$ fields. Completely arbitrary flavor indices are instead considered for the quark fields. Operators that violate baryon number are excluded because of the strong existing constraints. 

\begin{table}[t]
\begin{centering}
\begin{tabular}{|c|c|c|c|c|c|}
\hline 
\multicolumn{2}{|c|}{Four-lepton} & \multicolumn{4}{c|}{Semi-leptonic}\tabularnewline
\hline 
$\mathcal{O}_{\ldoublet\ldoublet}$ & $(\overline{\ldoublet}\gamma^{\mu}\ldoublet)(\overline{\ldoublet}\gamma_{\mu}\ldoublet)$ & $\mathcal{O}_{\ldoublet \qdoublet}^{(1)}$ & $(\overline{\ldoublet}\gamma^{\mu}\ldoublet)(\overline{\qdoublet}\gamma_{\mu}\qdoublet)$
& $\mathcal{O}_{\ldoublet u}$ & $(\overline{\ldoublet}\gamma^{\mu}\ldoublet)(\overline{u}\gamma_{\mu}u)$
\tabularnewline
 $\mathcal{O}_{ee}$ & $(\overline{e}\gamma^{\mu}e)(\overline{e}\gamma_{\mu}e)$ & $\mathcal{O}_{\ldoublet \qdoublet}^{(3)}$ & $(\overline{\ldoublet}\gamma^{\mu}\sigma^{I}\ldoublet)(\overline{\qdoublet}\gamma_{\mu}\sigma^{I}\qdoublet)$
& $\mathcal{O}_{\ldoublet d}$ & $(\overline{\ldoublet}\gamma^{\mu}\ldoublet)(\overline{d}\gamma_{\mu}d)$
 \tabularnewline
 $\mathcal{O}_{\ldoublet e}$ & $(\overline{\ldoublet}\gamma^{\mu}\ldoublet)(\overline{e}\gamma_{\mu}e)$ & $\mathcal{O}_{eu}$ & $(\overline{e}\gamma^{\mu}e)(\overline{u}\gamma_{\mu}u)$
& $\mathcal{O}_{\ldoublet ed\qdoublet }$ & $(\overline{\ldoublet}e)(\overline{d}\qdoublet )$
 \tabularnewline
&  & $\mathcal{O}_{ed}$ & $(\overline{e}\gamma^{\mu}e)(\overline{d}\gamma_{\mu}d)$
& $\mathcal{O}_{\ldoublet e\qdoublet u}^{(1)}$ & $(\overline{\ldoublet}e)\epsilon(\overline{\qdoublet }u)$
\tabularnewline
&  & $\mathcal{O}_{\qdoublet e}$ & $(\overline{\qdoublet}\gamma^{\mu}q)(\overline{e}\gamma_{\mu}e)$
& $\mathcal{O}_{\ldoublet e\qdoublet u}^{(3)}$ & $(\overline{\ldoublet}\sigma^{\mu\nu}e)\epsilon(\overline{\qdoublet }\sigma_{\mu\nu}u)$
\tabularnewline
\hline 
\end{tabular}
\par\end{centering}
\caption{Four-lepton and semi-leptonic baryon number conserving operators in the SM EFT.}
\label{tab:SMEFT_operators}
\end{table}

The Wilson coefficient $C_i$ of each interaction operator ${\mathcal{O}}_i$ is a parameter of energy dimension minus two, with an associated energy scale $\Lambda_i = |C_i|^{-1/2}$ that we often employ to report sensitivity projections.\footnote{If the boundaries of a confidence interval map to different values of $\Lambda$, we quote the smallest one.}
Notice that $\Lambda$ is only loosely related to the mass, $M_{\rm BSM}$, of the heavy BSM particles that are ultimately responsible for the emergence of the interaction in the low-energy EFT. The relation between the two quantities depends on the coupling of the BSM sector, on possible loop factors and/or symmetry selection rules and the only robust connection is the generic upper bound $M_{\rm BSM}\lesssim{\textrm{few}}\times\Lambda$, from the requirement of perturbative unitarity. Therefore, the $\Lambda\simeq100$\,TeV interaction scale sensitivity of the 10\,TeV MuC does not necessarily correspond to $M_{\rm BSM}=100$\,TeV BSM particle mass sensitivity. The BSM mass that is probed can be higher, if the BSM sector is strongly-coupled, or lower in the presence of a small coupling or of loop suppression. Our results apply as long as $M_{\rm BSM}>10$\,TeV because they are obtained assuming the validity of the EFT. For lower mass the reach must be assessed in the underlying BSM model exploiting also the possible direct production of the new particles.

\begin{figure}[t]
\centering
\includegraphics[width=0.45\textwidth]{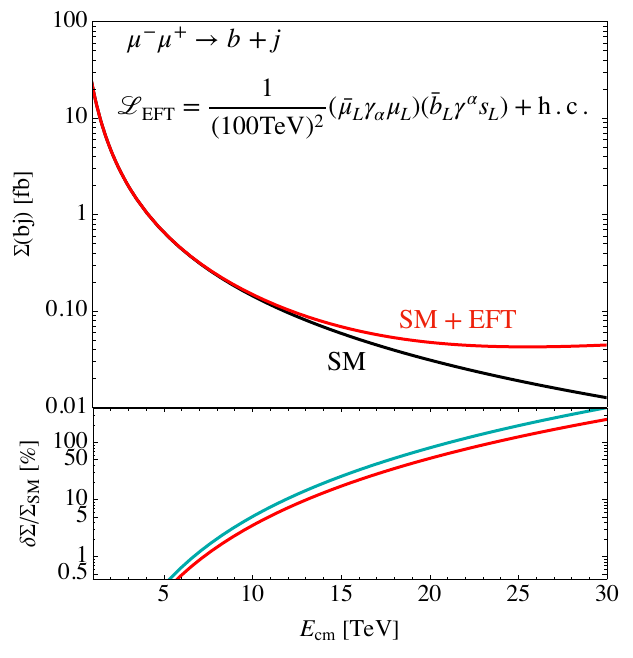}\quad    \includegraphics[width=0.46\textwidth]{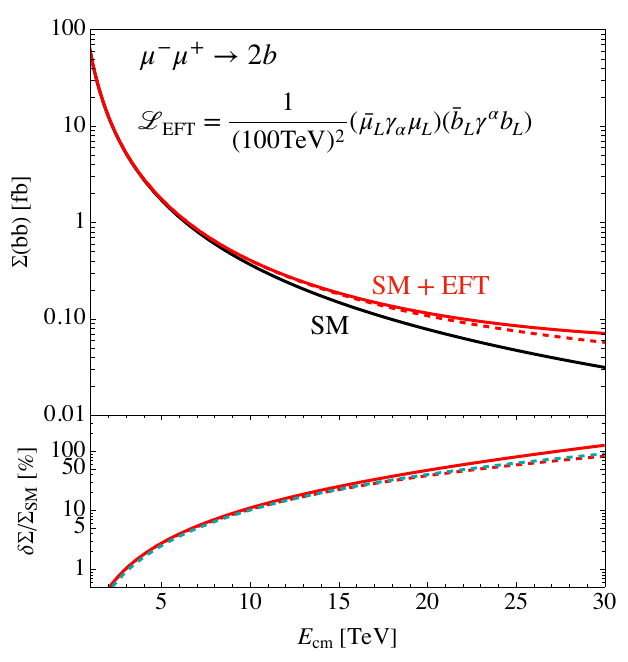}
    \caption{
    Exclusive cross sections for the $bj$ (left) and $bb$ (right) flavor-tagged dijet final states as a function of the collider center of mass energy. The black lines display the SM prediction, while the EFT interaction effects are reported in red for $\Lambda=100$\,TeV interaction scale. The bottom panels show the relative deviations from the SM due to the effective operator, in comparison with the estimates in Eqs.~(\ref{eq:estimate_FV_EFT0}) and~(\ref{eq:sigma_EFTeffect}) (blue). On the right panel, the EFT prediction excluding the quadratic terms are shown as a red dotted line.}
    \label{fig:sigma_mumubb}
\end{figure}

The Introduction provides a parametric estimate of the EFT scale sensitivity by exploiting the enhancement with energy of the new physics effect. The estimate in Eq.~(\ref{eq:estimate_FV_EFT0}) is for a flavor-changing interaction that mediates a process that is very rare in the SM and emerges from fakes with a fake probability ${\epsilon_{\rm f}}$. As a concrete illustration, on the left panel of Figure~\ref{fig:sigma_mumubb} we show the exclusive cross section for the $b+j$ diquark tagging category, as a function of the collider energy, obtained using  Eq.~\eqref{eq:x_seg_tag} and the exclusive cross section formula~\eqref{eq:xsec_general} integrated over in the angular region $\theta  \in [20^{\circ}, 160^{\circ}]$. The effects of a bottom-strange semi-leptonic EFT operator with $\Lambda_{\textrm{FC}}=100$\,TeV is also shown in the figure. The bottom panel reports the relative deviation from the SM due to the EFT operator.

In the SM, the $b+j$ state emerges from the production of a same-flavor quark-antiquark pair, of which either one or both are mistagged. Specifically:
$\Sigma(b+j)_{\rm SM} \approx 7\% \, \sigma(c\bar{c}) + 10\% \, \sigma(b\bar{b}) + 2\% \, \sum_{q=s,d,u}\sigma(q\bar{q})$. At the 10\,TeV MuC the SM cross sections are $\sigma(c\bar{c}) = 0.94\fb$, $\sigma(b\bar{b}) = 0.40\fb$, and $\sum_{q=s,d,u}\sigma(q\bar{q}) = 1.75\fb$. A large contribution is from $c\bar{c}$ production, even if it requires that both quarks are mistagged, due to the rather large mistag rates $\epsilon_c^j = 0.36$ and $\epsilon_c^b = 0.1$. Given the simultaneous contribution of different final states, the $b+j$ SM cross section can be estimated by an effective fake rate ${\epsilon_{\rm f}}=20\%$. The estimate~(\ref{eq:estimate_FV_EFT0}) is compared with the complete prediction in the bottom panel of the left panel of Figure~\ref{fig:sigma_mumubb}, showing agreement at order one.

With 10~ab$^{-1}$ (1~ab$^{-1}$) integrated luminosity, the cross section measurement in the $b+j$ final state has a statistical precision of $2.7\%$ ($2.2\%$) at the 10\,TeV (3\,TeV) MuC. The precision is nearly independent of the collider energy, owing to the quadratic scaling of the luminosity with energy, up to corrections from the Sudakov logarithms. The EFT effects are of order few percent at the 10\,TeV MuC and they reach the $1\sigma$ measurement sensitivity. The 10\,TeV MuC can thus probe the 100\,TeV interaction scale as expected from the estimate in Eq.~(\ref{eq:estimate_FV_SENS}). The EFT effects are much below percent at the 3\,TeV collider and the maximal scale that can be probed is about 30\,TeV.

Besides rare flavor-changing transitions, violations of unversality in the quark or lepton sector are other interesting probes of BSM flavor symmetry breakdown. The corresponding interactions preserve the flavor and contribute to final states that are unsuppressed in the SM. One such interaction, mediating the production of two b-quarks, is considered on the right panel of Figure~\ref{fig:sigma_mumubb}. The operator contributes to the $2b$ tagged category, which is well populated in the SM with a cross section of about $0.3$~fb at 10\,TeV ($4$~fb at 3\,TeV). The EFT contribution from a generic flavor-preserving interaction with scale $\Lambda_{\rm FP}$ can be estimated as follows. At the amplitude level, the EFT effect is of order $E_{\rm{cm}}^2/{\Lambda_{\rm FP}^2}$, to be compared with the SM amplitude that is of order $e^2\sim1/10$. The relative contribution to the cross section from the SM-EFT interference and the quadratic EFT term read, respectively:
\be\begin{split}
    \frac{\delta \sigma^{\rm int}_{\rm FP}}{\sigma_{\rm SM}} &\sim
        \frac{E_{\rm cm}^2}{e^2 \Lambda_{\rm FP}^2}
        \approx 10\% \left(\frac{E_{\rm cm}}{10~\TeV}\right)^2 \left(\frac{100~\TeV}{\Lambda_{\rm FP}}\right)^2~, \\
    \frac{\delta \sigma^{\rm quad}_{\rm FP}}{\sigma_{\rm SM}} &\sim
        \frac{E_{\rm cm}^4}{e^4 \Lambda_{\rm FP}^4}
        \approx 1\% \left(\frac{E_{\rm cm}}{10~\TeV}\right)^4 \left(\frac{100~\TeV}{\Lambda_{\rm FP}}\right)^4~.
        \label{eq:sigma_EFTeffect}
\end{split}\ee
In the parameter region of interest in our analysis the interference term dominates for those operators---such as the one considered on the right panel of Figure~\ref{fig:sigma_mumubb}---with the same chirality structure of the SM amplitudes. For the others, the interference vanishes and the quadratic term is the leading EFT effect. Using Eq.~(\ref{eq:sigma_EFTeffect}) and comparing with the statistical precision of about $1\%$ on the cros section measurement we obtain the estimates reported in Eq.~(\ref{eq:estimate_FP_SENS}) depending on whether the interference is present or not.

\begin{table}[t]
    \centering
    \begin{tabular}{|c|c|c|c|| c |}
    \hline
        Observable & Measurement & $\Lambda_{\rm vector}^{1\sigma}$ & $\Lambda_{\rm scalar}^{1\sigma}$ & 2040 precision \\ \hline\hline
        $R_{B_s \to \mu \mu}$ & $0.97 \pm 0.11$ \cite{ParticleDataGroup:2024cfk} & $44\, \TeV$ & $245\, \TeV$ & $4.4\%$ \cite{Cerri:2018ypt} \\ \hline
        $R_{B^0 \to \mu \mu}$ & $0.58 \pm 0.68$ \cite{ATLAS:2020acx} & $19\, \TeV$ & $145\, \TeV$ & $9.4\%$ \cite{Cerri:2018ypt} \\ \hline
        $R^{\nu}_{K^+}$ & $2.93 \pm 0.90$ \cite{Belle-II:2023esi} & $8 \, \TeV$ & -- & $11\%$ \cite{Belle-II:2018jsg} \\ \hline
        $\Br(D^0 \hspace{-3pt}\to\hspace{-2pt} \mu \mu)$ & $< 2.4 \times 10^{-9}, \, 95\%{\rm CL}$ \cite{CMS:2025fmx} & $8 \, \TeV$ & $40 \, \TeV$ & $1.8 \times 10^{-10}$ \cite{Cerri:2018ypt} \\ \hline
        $\Br(D^+ \hspace{-3pt}\to\hspace{-2pt} \pi^+ \mu \mu)$ & $< 8.0 \times 10^{-8}, \, 95\%{\rm CL}$ \cite{LHCb:2020car} & $8 \, \TeV$ & $8 \, \TeV$ & $3.7 \hspace{-2pt}\times\hspace{-2pt} 10^{-9}$ \cite{Cerri:2018ypt} \\ \hline
        $\Br(K_L \hspace{-3pt}\to\hspace{-2pt} \mu \mu)_{\rm SD}$ & $< 2.5 \times 10^{-9}$ \cite{Isidori:2003ts} & $80 \, \TeV$ & $370 \, \TeV$ & -- \\ \hline
        $R_{K \to \pi \nu \nu}$ & $1.57 \pm 0.36$ \cite{Buras:2015qea,NA62:2024pjp} & $90 \, \TeV$ & -- & 15\% \cite{HIKE:2023ext} \\ \hline
        $t \to \mu^- \mu^+ c (u)$ & $\Br(t \hspace{-3pt}\to\hspace{-2pt} Z q) < 2 \times 10^{-4}$ \cite{ATLAS:2018zsq} & $0.3 \, \TeV$ & $0.2 \, \TeV$ & $\sim 3 \times 10^{-5}$ \cite{Cerri:2018ypt} \\ \hline\hline
        $\Br(\tau^- \hspace{-3pt}\to\hspace{-2pt} \mu^- \mu^+ \mu^-)$ & $< 2.3 \times 10^{-8}, \, 95\%{\rm CL}$ \cite{Belle-II:2024sce} & $13 \, \TeV$ & $13 \, \TeV$ &  $3.3 \hspace{-2pt}\times\hspace{-2pt} 10^{-10}$ \cite{Belle-II:2018jsg} \\ \hline
        $\Br(\tau^- \hspace{-3pt}\to\hspace{-2pt} e^- \mu^+ \mu^-)$ & $< 3.2 \times 10^{-8}, \, 95\%{\rm CL}$ \cite{Hayasaka:2010np} & $10 \, \TeV$ & $10 \, \TeV$ & $4 \hspace{-2pt}\times\hspace{-2pt} 10^{-10}$\cite{Belle-II:2018jsg,Cerri:2018ypt} \\ \hline
        $\tau$ LFU & $\mathcal{O}(0.1\%)$ \cite{HeavyFlavorAveragingGroupHFLAV:2024ctg} & $8 \, \TeV$ & $1 \, \TeV$ & \\ \hline
        $e^+ e^- \to \mu^+ \mu^-$ & {\small LEP2,} & \multirow{2}{*}{$3.5 \, \TeV$} & \multirow{2}{*}{--} & \multirow{2}{*}{--} \\ 
        \cline{1-1}
        $\overset{\scriptscriptstyle(-)}{\nu}_\mu e^- \to \overset{\scriptscriptstyle(-)}{\nu}_\mu e^-$ & {\small CHARM(-II), BNL-E734} \cite{Falkowski:2015krw} & \multicolumn{1}{c|}{} & \multicolumn{1}{c||}{} & \multicolumn{1}{c|}{} \\ \hline
    \end{tabular}
    \caption{The low-energy flavor observables that dominate present-day constraints on the operators considered in the analysis. See Appendix~\ref{app:lowEflavor} for details. The third and fourth columns report the typical EFT scale presently constrained at 1$\sigma$ from each observable, for vector or scalar operators. The last column is the sensitivity reach achievable by the end of HL-LHC and Belle-II, circa 2040s.}
    \label{tab:lowenergy_obs}
\end{table}

\begin{figure}[t]
\centering
    \includegraphics[width=0.95\textwidth]{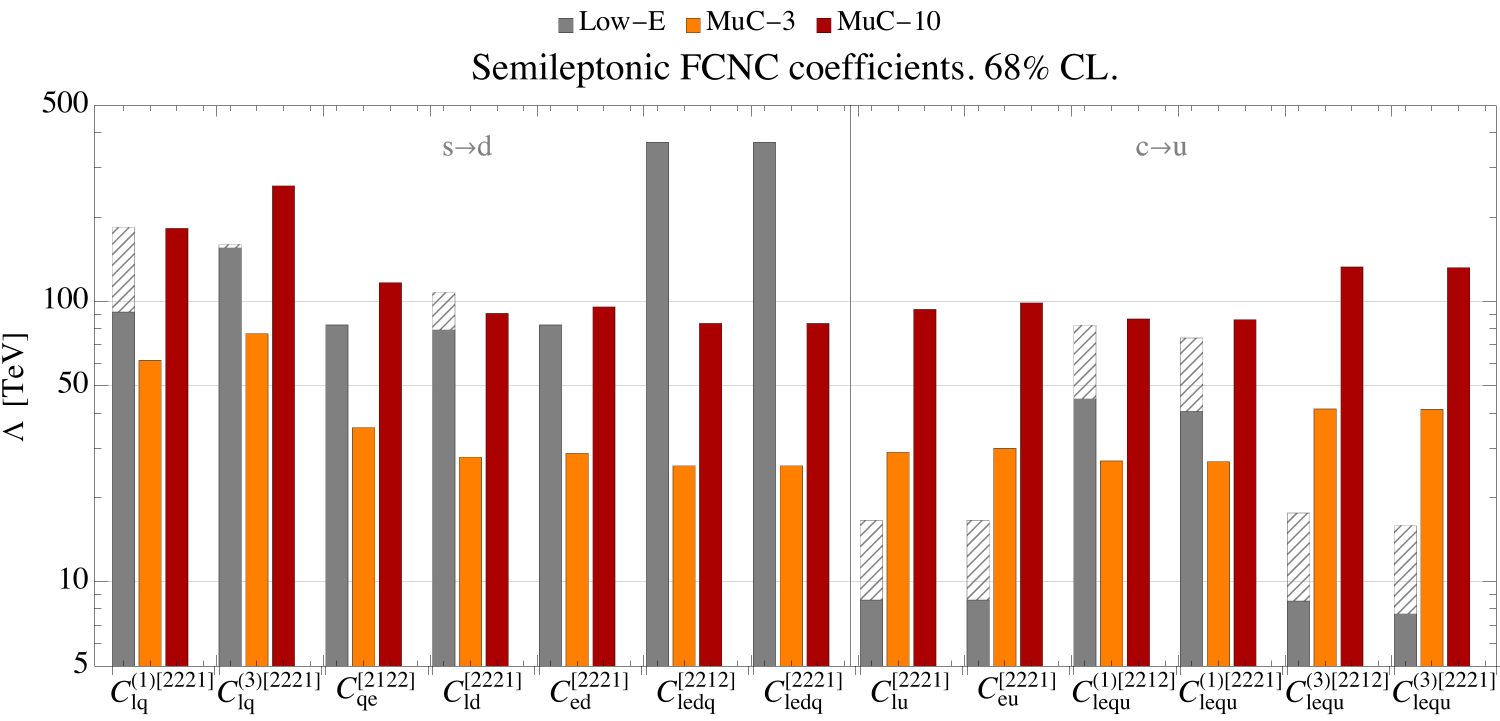}\\[0pt]
    \includegraphics[width=0.95\textwidth]{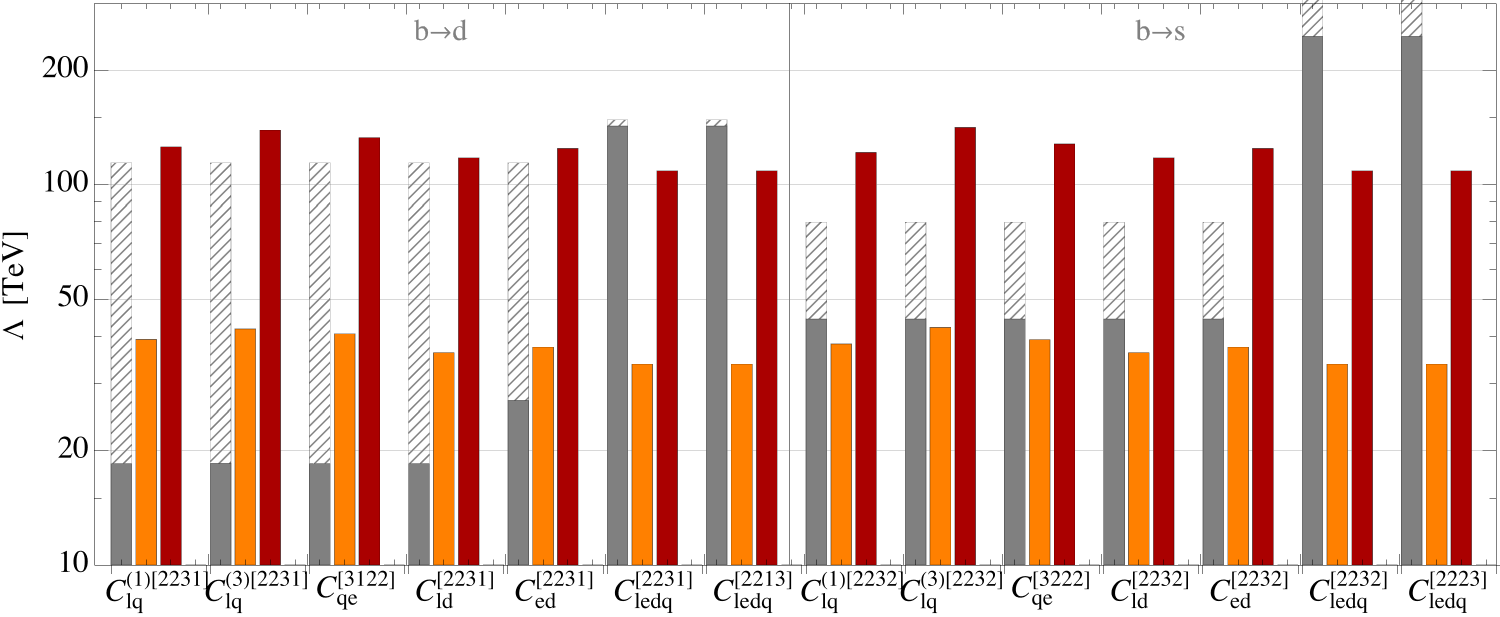}\\[0pt]
    \includegraphics[width=0.95\textwidth]{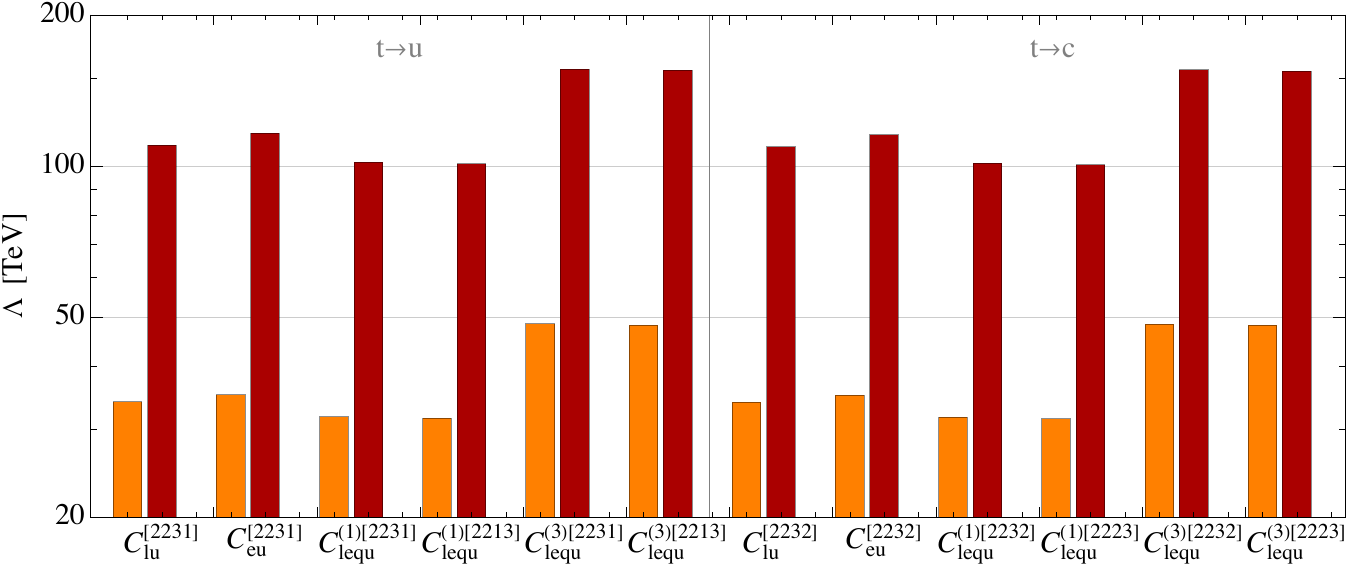}
    \caption{Single-operator sensitivity---namely, the sensitivity when all other operators are set to zero---at the 68\% CL from the 3\,TeV (orange) and 10\,TeV (red) MuC on semileptonic flavor-violating EFT coefficients, compared to present limits from low-energy measurements (gray). The hashed gray bars correspond to the expected sensitivity in the 2040s, with full HL-LHC, Belle-II, and NA62 results.}
    \label{fig:quarkFV_1D}
\end{figure}

The rest of this section illustrates the results of a likelihood fit to the operators in Table~\ref{tab:SMEFT_operators}, based on the differential cross section measurements in the diquark and dilepton final states detailed in Section.~\ref{sec:dif}. The projected reach of the 10\,TeV and of the 3\,TeV MuC is compared (when applicable) with existing constraints from low-energy observables. In Table~\ref{tab:lowenergy_obs} we summarize the most sensitive---see Appendix~\ref{app:lowEflavor}---low-energy probes of the contact interactions we study, together with an estimate of the typical EFT scales $\Lambda$ they are sensitive to. In the estimate we differentiate between operators constructed from vector or scalar currents, since the sensitivity can differ substantially in the two cases. The perspectives for improvements of the low-energy measurements by the HL-LHC and Belle-II are reported in the last column. 

The results are organized into subsections focusing on quark flavor-changing neutral currents, lepton flavor-violating interactions, and tests of flavor universality in both the quark and lepton sectors.

\subsection{Quark flavor-changing neutral currents}

The absence of tree-level contributions to FCNC processes in the Standard Model is a special accidental feature of the theory, and allows for very sensitive probes of flavor-violating new physics interactions. In this section we focus on effective semi-leptonic interactions---see Table~\ref{tab:SMEFT_operators}---that violate quark flavor and couple to two second-generation leptons.

\begin{figure}[t]
\centering
    \includegraphics[width=0.44\textwidth]{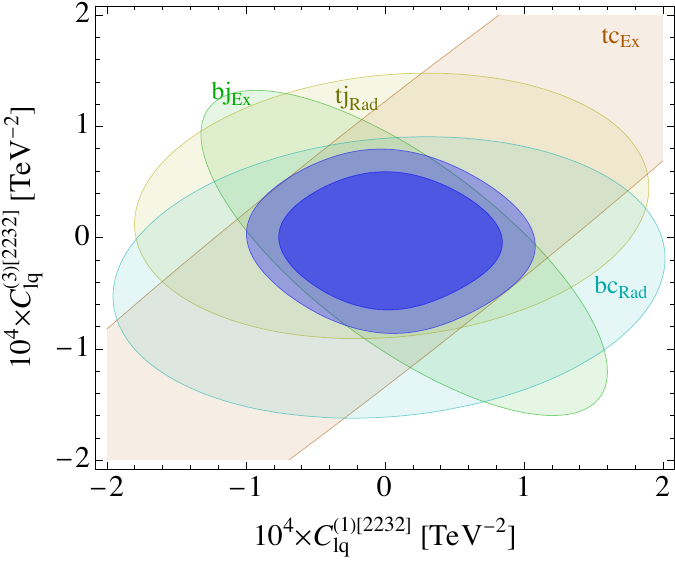} \quad
    \includegraphics[width=0.44\textwidth]{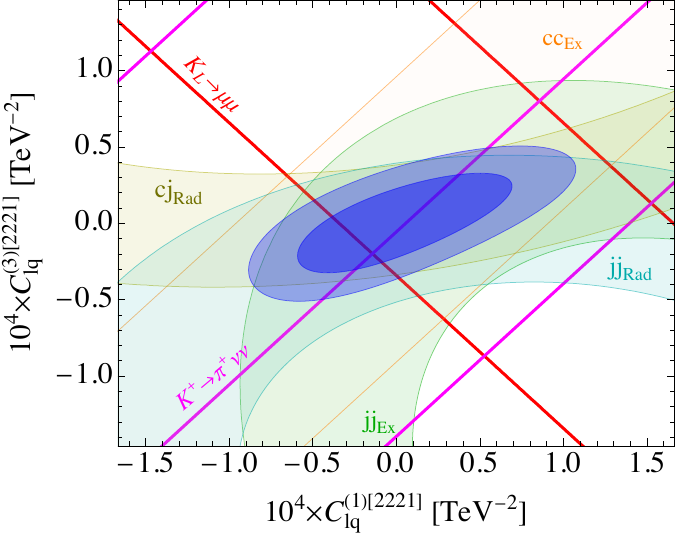}\\[0.5cm]
    \includegraphics[width=0.44\textwidth]{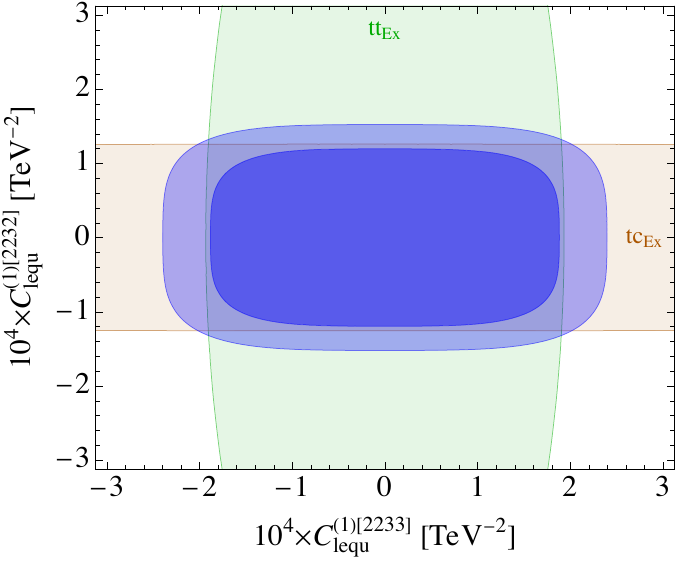}\quad
    \includegraphics[width=0.44\textwidth]{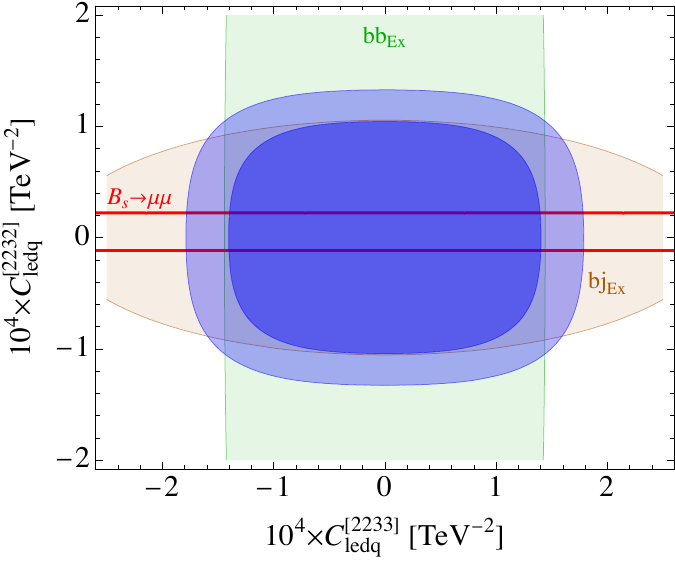}
    \caption{Illustrative fits of pairs of EFT coefficients at the 10\,TeV MuC. Dark and light blue are 68 and 95\% CL regions from the global analysis, respectively. Other contours are 68\% CL regions from individual flavor-tagged dijet categories, as indicated in the plots. The solid lines indicate current 68\% CL bounds from low-energy flavor observables.}
    \label{fig:dijet_2D}
\end{figure}

Figure~\ref{fig:quarkFV_1D} shows the single-operator sensitivity for all flavor-violating semi-leptonic EFT coefficients, comparing the reach of the 3\,TeV (orange) and 10\,TeV (red) muon collider with the one from current (gray) and future (hashed gray) low-energy flavor measurements. For scalar-current operators in the down sector, specifically $\mathcal{O}_{ledq}$ with all possible different family indices, the reach from leptonic meson decays surpasses the one of the 10\,TeV MuC. For all other cases, instead, the 10\,TeV muon collider is either marginally or substantially more sensitive than low-energy observables. Also, the 3\,TeV MuC offers a better or comparable reach in all cases except for $s \to d$ transitions. Substantial progress can be achieved by low-energy measurements, especially in the study of $b \to d$ and $b \to s$ transitions. Possible SM tensions resulting from these improvements could be clarified by the MuC probes of the same interactions in high-energy measurements. Besides matching or surpassing low-energy measurements, the muon collider also enables precise investigations of $t \to u$ and $t \to c$ transitions, currently constrained by top decays only at scales well below 1\,TeV.

Figure~\ref{fig:dijet_2D} displays 2-dimensional sensitivity contours on selected Wilson coefficient pairs. The global combined reach at the 10\,TeV MuC (blue) emerges from the combination of differential cross section (exclusive or with radiation) measurement in individual flavor-tagged dijet categories. The sensitivity from low-energy observables, when available, is reported as solid lines. The figure illustrates the MuC capability to disentangle different operators thanks to the availability of measurements in many different flavor-tagged categories. Degeneracies among operators with the same flavor structure are also addressed by the measurement of angular distributions and EW radiation, which are sensitive to the chirality and to the Lorentz structure of the operators.

Finally, in Table~\ref{tab:MuC_rareBdecays} we report the expected 68\%CL sensitivity on the individual coefficients of operators involving the third family of quarks, which can be probed at low energy in $B$ decays such as $B_s \to \mu\mu$, $B_d \to \mu\mu$ and $B^+ \to K^+ \nu\nu$. The table reports the relative precision in these rare $B$ decay branching ratios that would be needed to probe the coefficients at the same level as the 10\,TeV MuC. The present relative precisions are at the 10\%, 64\% and 90\% level for the $B_s \to \mu\mu$, $B_d \to \mu\mu$ and $B^+ \to K^+ \nu\nu$ branching ratios, respectively. The prospects for future measurements by the LHCb with detector upgrade-II and Belle-II with 50~ab$^{-1}$ of luminosity are approximately 4\%, 9\% and 11\%~\cite{Cerri:2018ypt,Belle-II:2018jsg}.

\subsection{Lepton flavor violation}
\label{sec:LFV}

We turn now to flavor-changing operators in the lepton sector, and compare the muon collider dilepton measurement sensitivity with the one of low-energy LVF searches. The relevant low-energy processes sensitive to the same operators---see Table~\ref{tab:lowenergy_obs}---are $\tau \to 3 \mu$ and $\tau \to e \mu^+ \mu^-$. A low-energy limit on $\mu \mu \mu e$ contact interactions can currently only be obtained indirectly, via loop contribution to $\mu \to e \gamma$, at the price of introducing model-dependency. A future possibility to test directly this interaction at low energy could come from LFV muonium decay $(\mu^- \mu^+) \to \mu^\pm e^\mp$ \cite{Minato:2025wcu}.

\begin{figure}[t]
\centering
    \includegraphics[width=0.98\textwidth]{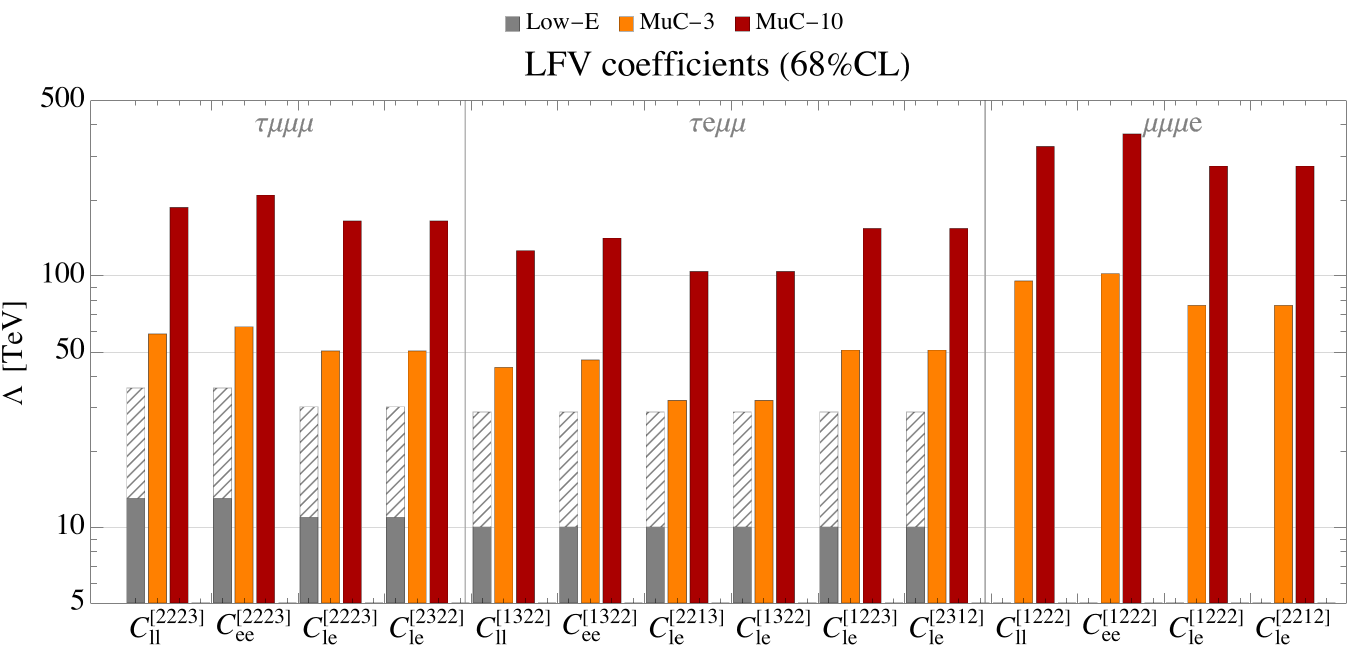}
    \caption{Single-operator sensitivity at the 68\% CL of the 10\,TeV (red) and 3\,TeV (orange)  MuC, and from $\tau$ decays (gray), on LFV EFT coefficients. The hashed gray bars correspond to the expected sensitivity with the full Belle-II dataset.}
    \label{fig:4lepton_LFV}
\end{figure}

We restrict our analysis to the exclusive cross section measurements in the $\ell^- \ell^{\prime +}$ (with $\ell \neq \ell^\prime$) final states. The same operators also contribute to with-radiation processes including the charged $\ell\nu$ final states and they might further improve the sensitivity. The results are reported in Figure~\ref{fig:4lepton_LFV} for the 3 and 10\,TeV MuC and compared with bounds from LFV $\tau$ decays. For a more quantitative comparison with the reach from low-energy observables, in Table~\ref{tab:MuC_tauLFV} we report the expected 68\% CL sensitivity of a 10\,TeV MuC on LFV SMEFT coefficients, together with the sensitivity on the LFV $\tau \to 3 \mu$ decay branching ratio measurement that would be needed to probe the operators at the same scale. Present experiments are sensitive to branching ratios at the $10^{-8}$ level.

\subsection{Testing flavor universality} 
\label{sec:flavor_universality}

In the limit where Yukawa couplings are neglected, the SM features an approximate global symmetry U$(3)^5$, since the three generations of the five fermionic irreducible SM representation have the same gauge charges. Equivalently, this is the statement that the SM gauge interactions are flavor universal. The SM Yukawa couplings break U$(3)^5$ in a peculiar manner, which preserves approximate universality in several interactions. For this reason, universality tests provide strong tests of new physics with the different pattern of flavor symmetry breakdown. Universality violation in the lepton sector is qualitatively different from that in the quark sector. Lepton Flavor Universality Violation (LFUV) and Quark Flavor Universality Violation (QFUV) are discussed below in turn.

\subsubsection*{Lepton flavor universality violation}

\begin{figure}[t]
\centering    \includegraphics[width=0.95\textwidth]{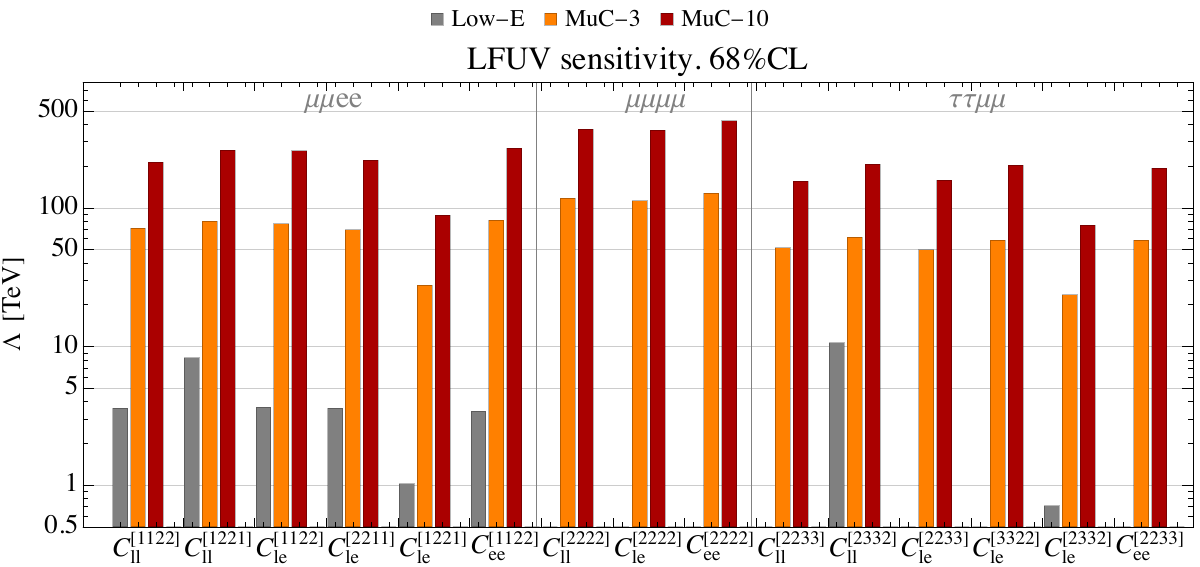}
    \caption{Single-operator sensitivity at the 68\% CL on flavor-preserving four-lepton operators with different lepton flavor of the 3\,TeV (orange) and 10\,TeV (red) MuC, compared to current constraints from low-energy measurements (gray).}
    \label{fig:4lepton_LFU}
\end{figure}

Figure~\ref{fig:4lepton_LFU} displays the single-operator sensitivity of our analysis for the 3 (orange) and 10\,TeV (red) MuC, which includes exclusive and with-radiation cross section measurements in the same-flavor dilepton final states $ee$, $\mu\mu$ and $\tau\tau$, and the charged $\ell\nu$ final state. These measurements can probe LFUV up to scales of $\sim 100 - 400 \, \TeV$ at a 10\,TeV~MuC, and $\sim 30 - 100 \, \TeV$ at the 3\,TeV collider.

Some of the strongest tests of LFUV at low energy can be derived from ratios of precisely measured $\tau$ decays~\cite{Stugu:1998jv,Pich:2013lsa}. 
The dependence of these ratios on the EFT operators is reported in Appendix~\ref{sec:tauLFU} and they provide sensitivity to scales of $\sim 1 - 10 \, \TeV$.  Among the operators contributing to these ratios, those that involve two muons {are $C_{ll}^{[1221]}$, $C_{le}^{[1221]}$, $C_{ll}^{[2332]}$, and $C_{le}^{[2332]}$.
Other observables sensitive to operators with two muons and two electrons are $\mu^+\mu^-$ production at LEP2 and muon (anti-)neutrino scattering on electrons, which have been analyzed in the EFT context by Ref.~\cite{Falkowski:2015krw} and probe scales of $\sim 3.5 \, \TeV$.
All these operators are very effectively probed at the muon collider, with a reach on the scale that is vastly superior to current and future knowledge even with the 3\,\TeV~MuC, as shown in Figure~\ref{fig:4lepton_LFU}.}

Another direct comparison with low-energy probes is presented in Table~\ref{tab:MuC_tauLFU}. We show the 95\% CL sensitivity reach of a 10\,TeV MuC on the corresponding Wilson coefficients, in comparison with the relative precision in $\tau$ LFU ratios that would be necessary to probe them at the same level. 
The last line of the table reports the reach for the operator $\mathcal{O}_{Hl}^{(3)} = (\bar l \gamma^\mu \sigma^I l)(H^\dagger \stackrel{\leftrightarrow}{D}_\mu \sigma^I H)$ from $\mu^+ \mu^- \to W W$~\cite{EWMuC}.
For all coefficients, a precision in $\tau$-decays LFU measurements at the level of $10^{-6}$ or much smaller would be needed in order to match the muon collider reach.

\begin{figure}[t]
\centering
    \includegraphics[width=0.44\textwidth]{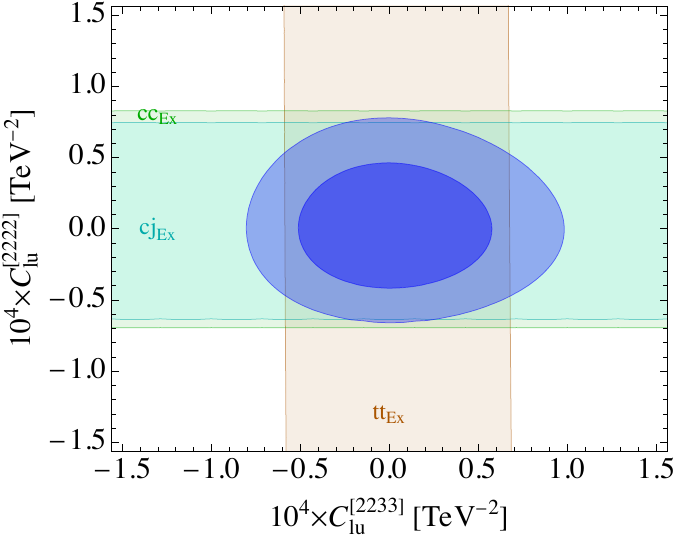} \quad
    \includegraphics[width=0.44\textwidth]{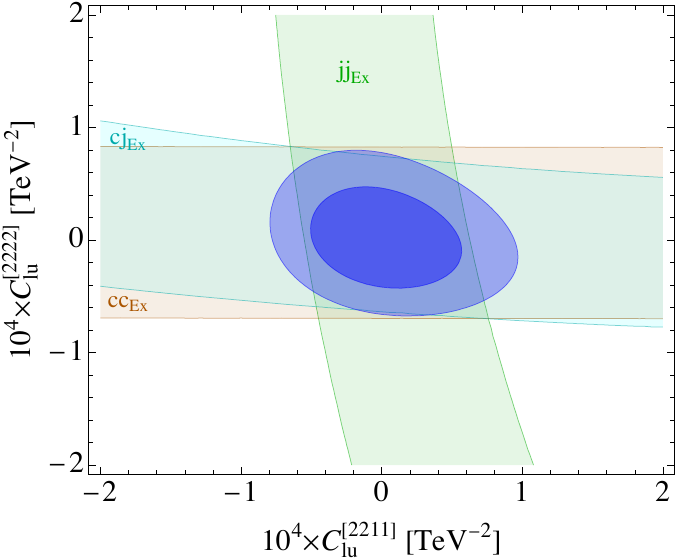}\\[0.5cm]
    \includegraphics[width=0.44\textwidth]{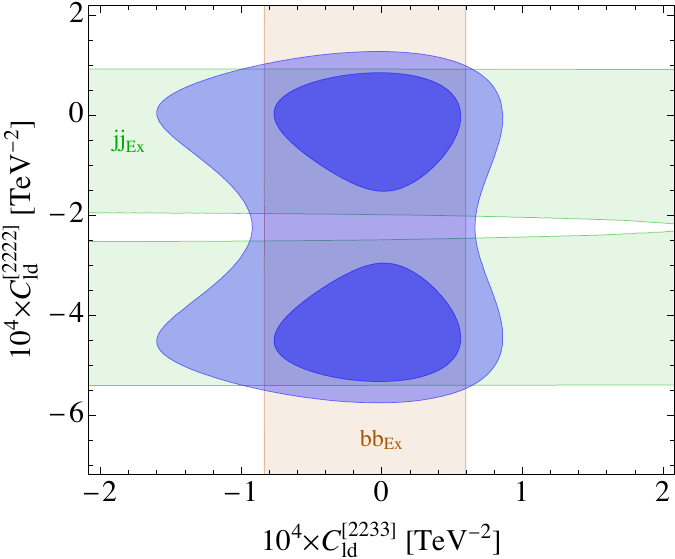} \quad
    \includegraphics[width=0.44\textwidth]{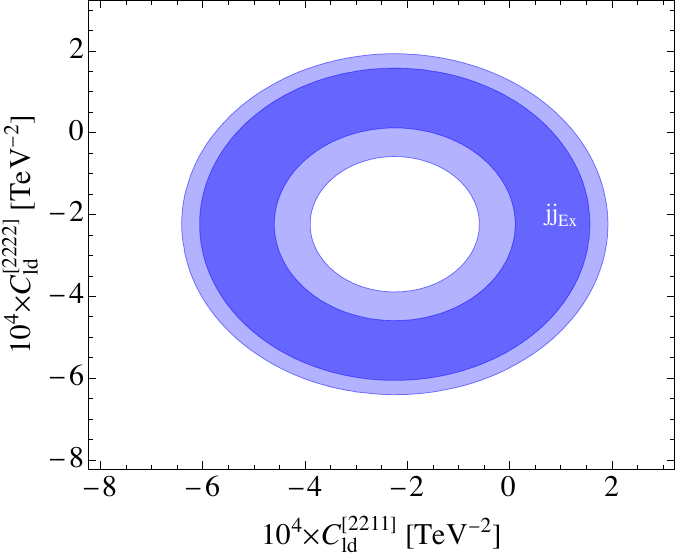}\\[0.5cm]
    \caption{Illustrative quark flavor universality tests for effective operators involving up (top) and down (bottom) quarks  at the 10\,TeV MuC. Dark and light blue are 68\% and 95\% CL regions, respectively. Other contours are 68\% CL regions from individual jet tagging categories, as indicated in the plots. The solid lines indicate current 68\% CL bounds from low-energy flavor observables.}
    \label{fig:dijet_QFU}
\end{figure}

\subsubsection*{Quark flavor universality violation}

In traditional low-energy experiments or at hadron colliders, QFU is not a good approximate symmetry, because of several reasons. First, the top quark Yukawa breaks QFU maximally and also the CKM mixing between the first two generations is large, with the Cabibbo angle being $\sin \theta_{C} \approx 0.2$. For low-energy observables, the squared mass of pseudo-scalar mesons is proportional to quark masses, therefore to the breaking of universality, and the decay phenomenology depends crucially on the valence quark flavors. Finally, in high-energy proton-proton colliders the proton PDF is also an $\mathcal{O}(1)$ source of QFUV. The situation is different at a muon collider, where the initial state carries no quark flavor charge and the masses of all quark masses (including the top) can be treated as small perturbations due to the large collision energy. Therefore, at a muon collider it becomes meaningful to ask what is the capability of testing the universality of new physics interactions among the three families, by using the discriminating power of the different jet flavor tagging categories.

Figure~\ref{fig:dijet_QFU} shows some representative examples of QFUV tests with effective operators involving up (top panels) or down (bottom panels) quarks. The good bottom and charm tagging, with low mistag rates from light quarks, and boosted-top identification allow to separately identify new physics interacting with the third quark generation from interactions with the first two (top left and bottom left panels), and to disentangle operators involving charm vs.\ up quarks (top right panel). On the other hand, our analysis is unable to differentiate interactions involving strange against down quarks (bottom right panel). This could be improved by implementing strange tagging.


The combined sensitivity to flavor-preserving and to flavor-changing operators enables the muon collider to probe the complete flavor structure of semi-muonic operators. As an example, consider the $C_{lq}^{(3)\,[22ij]}$ coefficients matrix---with $i, j$ indices for the quark flavors---and compare present and future low-energy reach at 68\% CL on the EFT scale $\Lambda$ with the muon collider projections:
\begin{equation}
    \begin{split}
    \Lambda_{ij} = \left|C_{lq}^{(3)\,[22ij]}\right|^{-1/2} ~(1\sigma):& ~ 
    \left( \begin{tabular}{ccc}
       --  & 155 & 18 \\
       155 & -- & 44 \\
       18 & 44 & --
    \end{tabular} \right)_{\text{low-E}}^{\rm today}
    \!\!\!\!\!\!\!\!\!\!\longrightarrow
    \left( \begin{tabular}{ccc}
       --  & 159 & 114 \\
       159 & -- & 80 \\
       114 & 80 & --
    \end{tabular} \right)_{\text{low-E}}^{\rm future} \\
    & ~ 
    \left( \begin{tabular}{ccc}
       411  & 259 & 139 \\
       259 & 397 & 141 \\
       139 & 141 & 431
    \end{tabular} \right)_{\text{MuC-10}}
\end{split}
\end{equation}
While only the $O_{lq}^{(3)}$ operator is considered for illustration, similar results are found for the other current-current semi-muonic operators. The 10\,TeV muon collider offers unparalleled perspective to probe the complete flavor structure of such new physics scenarios.

\section{Conclusions and outlook}
\label{sec:conclusions}

We presented an exploratory assessment of the flavor physics potential of high-energy muon colliders. This potential does not stem from improving the precision of traditional low-energy flavor physics measurements, but rather from the direct access to the flavor-breaking interactions in high-energy measurements. This defines a novel methodology for flavor physics exploration at the energy rather than at the intensity frontier, which can be leveraged also at the LHC or other future high energy colliders \cite{Cirigliano:2012ab,Murakami:2014tna,Faroughy:2016osc,Greljo:2017vvb,Greljo:2018tzh,Fuentes-Martin:2020lea,Angelescu:2020uug,Marzocca:2020ueu,Endo:2021lhi,Crivellin:2021rbf,Greljo:2022jac,Allwicher:2022gkm,Altmannshofer:2023tsa,Greljo:2024ytg,Altmannshofer:2025nbp} but with less competitive perspectives.

Our main result---in Section~\ref{sec:FC}---is the 100\,TeV scale sensitivity of the 10\,TeV muon collider to a variety of four-fermion operators---specifically, all those that contain a muon bilinear $(\bar\mu \Gamma \mu)$---that break flavor symmetries in the quark or in the lepton sector. The 100\,TeV sensitivity is attained both for operators inducing rare flavor-changing transitions and for flavor-preserving operators that violate flavor universality. Some of the tested interactions---for instance the ones involving the top quark---are not accessible with low-energy measurements. On the others, the muon collider sensitivity reach is competitive and in some cases strongly superior to the one that is achievable with low-energy measurements, also taking into account the prospects for improvement during the next decade. The sensitivity of a possible 3\,TeV stage of the muon collider, at the 30\,TeV scale, is still very competitive for some EFT interactions.

The high-energy muon collider probes are interesting not only for their high scale sensitivity. Unlike traditional low-energy flavor physics probes, the muon collider measurements are less exposed to experimental or theoretical mismodeling uncertainties as they do not target the observation of extremely rare phenomena, nor they rely on extremely accurate measurements and theoretical predictions. They exploit the large available energy that mitigates the suppression of the EFT operators enhancing their effects. The presence of new physics would produce a relatively large (above $\%$) departure from the SM predictions, emerging in multiple final states and correlated in different angular bins. These observations would provide a convincing evidence of new physics and they would enable a rather accurate characterization of the discovery. Similar considerations apply if the first observed tension with the SM arises from low-energy measurements: muon collider measurements would be key for characterization. 

\subsection*{Directions for future work}

Our assessment of the muon collider flavor physics opportunities could be expanded and refined in multiple directions. First, by the study of explicit models offering a complete description of the BSM particles dynamics. The EFT offers a complete model-independent description, but only if the BSM particles mass is above the collider energy, namely if $M_{\rm BSM}$ is larger than 10\,TeV, for the 10\,TeV collider. If $M_{\rm BSM}$ is smaller, the flavor breaking effects are generically larger but the sensitivity reach cannot be extrapolated from our results. On the one hand, our sensitivity emerges from the quadratic growth with energy of the EFT amplitude, which matches the true behavior of the underlying complete BSM model only at energies below $M_{\rm BSM}$. On the other hand, if $M_{\rm BSM}$ is below the collider energy the search for resonant new particles production can improve the sensitivity to the BSM model.

Significant improvements are also needed on the theoretical accuracy of the high-energy cross section predictions. Even if the target accuracy is only at the percent level, attaining this accuracy requires more than the resummation of the EW Sudakov double logarithms that we account for in our predictions. Single logarithms should be definitely included and possibly resummed. The matching with complete fixed-order calculations might be also needed to reach the required accuracy. A significant literature exists---see the overviews in~\cite{Accettura:2023ked,InternationalMuonCollider:2025sys}---on the treatment of IR-enhanced EW effects, but the feasibility of the predictions at the required level of accuracy has not been demonstrated yet. See Refs.~\cite{Ma:2024ayr,Frixione:2025guf,Frixione:2025wsv} for recent progress on EW corrections for muon collider processes.

Our results are based on a crude treatment of the experimental effects. This is unavoidable, given the yet preliminary stage of the muon collider experiments design and the difficulty of extrapolating to the new conditions the performances of better known detectors at the LHC or at future low-energy lepton colliders. We believe that our scale sensitivity projections are robust, because they do not rely on aggressive flavor-tagging performances or on the strong reduction of otherwise overwhelming backgrounds. Still, detailed experimental work is needed to put our findings on firmer grounds.

\section*{Acknowledgments}

This work has been endorsed by the IMCC. We acknowledge valuable feedback on the draft by the IMCC Publication Speakers Committee and in particular M.~Loeschner and F.~Meloni. The work started in the context of the preparation of the input to the 2026 update of the European Strategy for Particle Physics and some of its results are documented in the IMCC submission~\cite{InternationalMuonCollider:2025sys}. We thank J.~de Blas, R.~Franceschini and X.~Wang for collaboration at the initial stage of the work. We also acknowledge for hospitality the Galileo Galilei Institute for Theoretical Physics (GGI) where part of this work was completed during the GGI Workshop ``Exploring the energy frontier with muon beams''.

The work of AG was supported in part by the European Union - Next Generation EU under Italian MUR grant PRIN-2022-RXEZCJ.
DM acknowledges support from the European Union - Next Generation EU under Italian MUR grant PRIN~20224JR28W. 
AW acknowledges financial support from the Spanish Ministry of Science and Innovation (MICINN) through the Spanish State Research Agency, under Severo Ochoa Centres of Excellence Programme 2025-2029 (CEX2024001442-S). 
This work is part of the R\&D\&i project PID2023-146686NB-C31, funded by MICIU/AEI/10.13039/ 501100011033/ and by ERDF/EU.

\appendix

\section{Resummation of EW double logarithms}
\label{app:EWresum}

This Appendix briefly reviews the basic formulae for the resummation of EW Double Logarithms (DL). Their derivation and implementation for difermion and diboson processes can be found in~\cite{Chen:2022msz}, while a more detailed discussion of the CKM rotation implementation is described below. The predictions employed in the paper including EFT operator effects can be obtained with a publicly available code~\cite{Glioti:2025}.

The resummation of electroweak infrared effects is accomplished through evolution equations. These equations describe the evolution of physical quantities as an infrared cutoff $\lambda$ is lowered from the scattering energy scale $s=E^2$ to $m^2_W$. At the DL level, these equations can be computed and solved analytically for the two classes of observables discussed here.
 
The infrared evolution equations are correct only in the limit in which the weak SU$(2)_L$ symmetry is exact. This means that in our calculations for the hard Born matrix elements we will treat all particles as massless and we will ignore the contributions in the amplitude that are proportional to powers of $m_{\textrm{w}}/E$. Explicitly, this means that Standard Model amplitudes will only contain $E^0$ energy dependence while the BSM amplitudes from $d=6$ operators will only contain the $E^2$ terms. 

Single-logarithm effects can also be numerically important for multi-TeV lepton colliders and should be included or possibly resummed to achieve the (percent) level of accuracy on the theoretical predictions that would be needed for the comparison with the actual data. A much lower precision is sufficient for our sensitivity estimates and single-logarithm corrections can be neglected. However, these corrections become especially large in the forward/backwards region. For this reason, we restrict our analysis to the central angular region $\theta \in [20^\circ, 160^\circ]$.

\subsection*{Exclusive final state}

As explained in the main text, we define exclusive final state as the processes $\mu^- \mu^+ \to f\bar{f}$ with no soft electroweak radiation in the final state. The only effects we consider are the corrections coming from virtual exchanges of electroweak bosons between the external particles. 

At the DL level, the Sudakov effects on the scattering amplitudes can be resummed into a leg-by-leg exponential factor that only depends on the electroweak quantum number of the initial and final states. Explicitly, every squared scattering amplitude with a given helicity of the initial and final states is multiplied by Sudakov factors
\begin{equation}
    \prod_i S_i = \prod_i \exp\left( - \frac{g^2(E) c_i + g^{\prime 2}(E) Y_i^2}{16 \pi^2} \log^2 \frac{E^2}{m_{\textrm{w}}^2} \right)~,
    \label{eq:exclSudakov}
\end{equation}
where the index $i$ runs over the external particles, $c_i$ are the SU$(2)_L$ Casimir (equal to $3/4$ 
for doublets and to 0 for singlets), $Y_i$ are the hypercharges, and $g(E)$, $g^{\prime}(E)$ the SU$(2)_L$ and U$(1)_Y$ gauge couplings evaluated at the scale $E$. The exclusive cross sections with polarized external states are then found as
\begin{equation}
    \frac{d \sigma_\text{exc}(f \bar f^\prime)}{d \cos\theta} = \left(\prod_i S_i\right) \frac{d \sigma_{\rm Born}(f \bar f^\prime)}{d \cos\theta}~.
    \label{eq:xsec_general}
\end{equation}

\subsection*{Semi-inclusive final state}

For the semi-inclusive final state, that is $\mu^- \mu^+ \to f\bar{f} + X$, we instead take into account all virtual and real electroweak radiation. Notice that the radiation is soft (and collinear) at the DL level. Therefore we can neglect the energy loss due to radiation and assume that the hard scattering happens at the full collider center of mass energy $E$, up to single log corrections.

Since the emission of a $W^\pm$ from either the initial or final legs changes the particles that participate in the hard scattering, several scattering amplitudes with different external legs are simultaneously relevant for the semi-inclusive cross section calculation. For this reason, the semi-inclusive DL resummation cannot be performed at the amplitude level, but rather at the level of the density matrix. Including flux and phase-space factors (and integrating over the azimuthal angle) the Born-level density matrix is defined as
\begin{equation}
\label{eq:dmB}
\mathcal{D}^{\alpha_1\dots\alpha_4 \bar\alpha_1 \dots \bar\alpha_4}_\text{Born} \equiv \frac{1}{32\pi E^2} \mathcal{M}_\text{Born}^{\alpha_1\dots\alpha_4 }\left[\mathcal{M}_\text{Born}^{\bar\alpha_1 \dots \bar\alpha_4}\right]^* \,,
\end{equation}
where the $\alpha_i$ index is the SU$(2)_L$ index of the $i$ external particle and $\mathcal{M}_\text{Born}$ is the Born-level scattering amplitude. 

The resummed semi-inclusive density matrix is then given by the leg-by-leg mixing
\begin{equation}
\label{eq:resSI0}    
\mathcal{D}^{\alpha_1\dots\alpha_4 \bar\alpha_1 \dots \bar\alpha_4}_\text{SI} = [\mathcal{K}_1]^{\alpha_1 \bar\alpha_1}_{\beta_1 \bar\beta_1} \dots [\mathcal{K}_4]^{\alpha_4 \bar\alpha_4}_{\beta_4 \bar\beta_4} \mathcal{D}^{\beta_1\dots\beta_4 \bar\beta_1 \dots \bar\beta_4}_\text{Born} \,,
\end{equation}
where the tensor $\mathcal{K}_i$ relative to each external leg ``$i$'' is trivially equal to $1$, for SU$(2)$ singlets, and 
\begin{equation}
    [\mathcal{K}]^{\alpha \bar\alpha}_{\beta \bar\beta} = e^{-\mathcal{L}} \delta^{\alpha}_\beta \delta^{\bar\alpha}_{\bar\beta} + e^{-\mathcal{L}/2} \sinh (\mathcal{L}/2) \delta^{\alpha \bar\alpha}\delta_{\beta \bar\beta} \,,
\end{equation}
for SU$(2)$ doublets, where
\begin{equation}
    \mathcal{L} = \frac{g(E)^2}{16\pi^2} \log^2 \left( \frac{E^2}{m_{\textrm{w}}^2} \right) \,.
\end{equation}

It should be noted that the scattering amplitudes and the density matrix are defined for gauge eigenstate external states. In the case of leptons, these coincide with the mass eigenstates of definite helicity, while in the case of quarks the CKM matrix rotation must be taken into account. A simple explicit formula is thus obtained for dileptons while the expression is complicated for diquark production, as discussed below in turn.

\subsubsection*{Dilepton production}

For each lepton family, the gauge eigenstate particles consist of two singlets, i.e.\ the helicity-plus charged lepton $e_+$ and the helicity-minus anti-lepton $\bar{e}_-$, one doublet $l^\alpha=(\nu_-,e_-)$ with negative helicity and one anti-doublet $\bar{l}^\alpha=(\bar{\nu}_+,\bar{e}_+)$. Eq.~(\ref{eq:resSI0}) can be applied directly, where legs 1 and 2 are second-generation lepton doublets or singlets depending on the polarization of the incoming beam particles, and legs 3 and 4 are the final-state lepton and anti-lepton with arbitrary lepton flavor.

The scattering cross sections are equal to the diagonal entries of the density matrix, namely to the entries with $\bar\alpha_{i}=\alpha_{i}$. When $\bar\alpha=\alpha$, the kernel tensors $\mathcal{K}_i$ simplify and become diagonal in $(\beta, \bar\beta)$, namely
\begin{equation}
    [\mathcal{K}]^{\alpha \alpha}_{\beta \bar\beta}
    =\mathcal{S}^\alpha_\beta \delta_{\beta\bar\beta}\,,
\end{equation}
where $\mathcal{S}=1$ in the case of singlets while for doublets 
\begin{equation}
    \mathcal{S}^\alpha_\beta=
 \left[
    \begin{array}{cc}
    e^{-\mathcal{L}/2} \cosh (\mathcal{L}/2) & e^{-\mathcal{L}/2} \sinh (\mathcal{L}/2)
        \\
    e^{-\mathcal{L}/2} \sinh (\mathcal{L}/2) & 
    e^{-\mathcal{L}/2} \cosh (\mathcal{L}/2) 
    \end{array}
    \right]_{\alpha\beta}.
\end{equation}
Using this in Eq.~(\ref{eq:resSI0}) we find that the differential cross section of the semi-inclusive processes can be expressed in terms of the corresponding Born-level cross sections. Namely
\begin{equation}
\label{eq:resSI}
{\textrm{for dilepton production:}}\qquad\frac{d\sigma_{\textrm{SI}}^{\alpha_1\dots\alpha_4}}{d\cos\theta}=
[\mathcal{S}_1]^{\alpha_1}_{\beta_1}
\dots 
[\mathcal{S}_4]^{\alpha_4}_{\beta_4}
\frac{d\sigma_{\textrm{Born}}^{\beta_1\dots\beta_4}}{d\cos\theta}\,.
\end{equation}
Unlike in the exclusive case~(\ref{eq:xsec_general}), the Sudakov factors are not diagonal in the SU$(2)_L$ indices. The semi-inclsive cross sections are a linear combination of Born cross sections with different external state. This enables for instance the production of a fermion pair with non-vanishing total charge out of the neutral $\mu^-\mu^+$ inital state.

\subsubsection*{Diquark production}

In the case of quarks, the right-handed helicity-plus mass eigenstates $u_+$ $d_+$ (and the anti-particles $\bar{u}_-$ and $\bar{d}_-$) are also gauge eigenstates, in the singlet representation. The other states are cast in a doublet and an anti-doublet, taking however into account the CKM rotation on the flavor indices. Namely, we have
\begin{eqnarray}
\label{eq:MGb}
    q^{\alpha;f}_{\textsc{m}}=
    \left[
    \begin{array}{r}
    u_-^f 
        \\
    d_-^f
    \end{array}
    \right]
    =
    \left[
    \begin{array}{r}
        V_{fg}q^{1;g} 
        \\
    \delta_{fg}q^{2;g}
    \end{array}
    \right]
    \equiv{\mathcal{V}}^{\alpha;f}_{\beta;g}q^{\beta;g}
    ,\nonumber\\
    {\bar{q}}^{\alpha;f}_{\textsc{m}}=
    \left[
    \begin{array}{r}
    u_-^f 
        \\
    d_-^f
    \end{array}
    \right]
    =
    \left[
    \begin{array}{r}
        V_{fg}^*q^{1;g} 
        \\
    \delta_{fg}q^{2;g}
    \end{array}
    \right]
    \equiv\bar{\mathcal{V}}^{\alpha;f}_{\beta;g}{\bar{q}}^{\beta;g}
    ,
\end{eqnarray}
where $q_{\textsc{m}}$ and ${\bar{q}}_{\textsc{m}}$ are mass-eigenstate doublets, $V$ is the CKM matrix and $q$, ${\bar{q}}$ are the gauge-eigenstate doublets. 

Given the interplay between gauge and flavor indices, a generalization of the density matrix needs to be considered for the DL resummation, taking into account the flavor indices of the final-state quarks. Namely, we define
\begin{equation}
\label{eq:genDM}
\mathcal{D}^{\alpha_1\alpha_2(\alpha_3g_3)(\alpha_4g_4) \bar\alpha_1 \bar\alpha_2(\bar\alpha_3\bar{g}_3) (\bar\alpha_4\bar{g}_4)
}_\text{Born} \equiv \frac{1}{32\pi E^2} \mathcal{M}_\text{Born}^{\alpha_1\alpha_2(\alpha_3g_3)(\alpha_4g_4)}\left[\mathcal{M}_\text{Born}^{\bar\alpha_1 \bar\alpha_2(\bar\alpha_3\bar{g}_3) (\bar\alpha_4\bar{g}_4)}\right]^* \,,
\end{equation}
where the Born amplitude $\mathcal{M}_\text{Born}$ depends explicitly on the quark flavor indices $g_{3,4}$ and ${\bar{g}}_{3,4}$. Notice that the Born amplitude is evaluated with gauge eigenstate external states, i.e.\ before the CKM rotation. The DL resummation of this generalized density matrix works as before, with the same kernel tensors as in Eq.~(\ref{eq:resSI0}) that only act on the gauge indices.

The generalized density matrix needs to be rotated to the physical basis of mass eigenstates using Eq.~(\ref{eq:MGb}), and the scattering cross sections are the diagonal entries $\bar\alpha_i=\alpha_i$ and $\bar{g}_i=g_i$ after the CKM rotation. The final expression is involved and not worth reporting here. Unlike for dileptons~(\ref{eq:resSI}), the semi-inclusive diquark cross section is not a linear combination of Born-level cross sections. Off-diagonal entries of the generalized density matrix~(\ref{eq:genDM}) appear in the expression.

\section{Low-energy flavor observables}
\label{app:lowEflavor}

This Appendix contains a brief description of the low-energy observables we use to compare with the muon collider EFT sensitivity reach.
For each flavor changing transition involving two muons or muon neutrinos, we select the most sensitive observable and report the corresponding experimental measurement and SM prediction.

\subsection*{$\boldsymbol{b \to s \mu \mu}$ and $\boldsymbol{b \to d \mu \mu$}}

There is a large number of observables sensitive to these flavor transitions, however most are not well suited to our analysis.
The angular distributions of $B \to K^* \mu \mu$, as well as branching fractions of other semi-leptonic rare $B$ decays, are plagued by long-distance non-pertubative QCD contributions, which are currently not under firm theoretical control. For this reason we do not consider them here.
On the other hand, the LFU tests in $R(K)$ and $R(K^*)$, which are theoretically clean and could offer strong sensitivity, require additional assumptions on semi-leptonic operators with electrons which instead do not affect our muon collider analysis.

For these reasons we consider only the leptonic rare decays $B_{s} \to \mu \mu$ and $B^0 \to \mu \mu$.
For $B_{s} \to \mu \mu$ we take the 2024 PDG average \cite{ParticleDataGroup:2024cfk} of experimental measurements, $\Br(B_s \to \mu \mu)_{\rm exp} = (3.52 \pm 0.32) \times 10^{-9}$, while for $B^0 \to \mu \mu$ we take the 2020 experimental average of ATLAS, CMS, and LHCb results \cite{ATLAS:2020acx}, $\Br(B^0 \to \mu \mu)_{\rm exp} = (0.6 \pm 0.7) \times 10^{-10}$.
We use the SM prediction for $B^0 \to \mu \mu$ from Ref.~\cite{Beneke:2019slt}: $\Br(B^0 \to \mu \mu)_{\rm SM} \approx 1.03 \times 10^{-10}$.
For $B_s \to \mu \mu$ we start with the recent analysis of Ref.~\cite{Fleischer:2024fkm}, taking as central value the average of the predictions obtained with the inclusive and exclusive $V_{cb}$ measurements, and half of their difference as uncertainty: $\Br(B_s \to \mu \mu)_{\rm SM} = (3.45 \pm 0.26) \times 10^{-9}$.

For convenience, we define the ratios of experimental measurement to SM prediction:
\be\begin{split}
    R_{B_s \to \mu \mu} &= \frac{\Br(B_s \to \mu \mu)_{\rm exp}}{\Br(B_s \to \mu \mu)_{\rm SM}} = 0.97 \pm 0.11~, \\
    R_{B_d \to \mu \mu} &= \frac{\Br(B^0 \to \mu \mu)_{\rm exp}}{\Br(B^0 \to \mu \mu)_{\rm SM}} = 0.58 \pm 0.68~.
\end{split}\ee
We take the EFT dependence from Ref.~\cite{Fleischer:2024fkm} and match the low-energy EFT with the SMEFT coefficients at tree-level. By the end of the HL-LHC run, the three LHC experiments are expected to reduce the uncertainties on the muonic $B_s$ and $B^0$ branching ratios to 4.4\% and 9.4\%, respectively \cite{Cerri:2018ypt}.

\subsection*{$\boldsymbol{c \to u \mu \mu}$}

This flavor transition is probed at low energy by the rare $D$-meson decays $D^0 \to \mu \mu$ and $D^+ \to \pi^+ \mu \mu$. The experimental upper limits on both are
\begin{equation}\begin{split}
    \Br(D^0 \to \mu \mu) <& 2.4 \times 10^{-9},\, ~ 95\%{\rm CL} \quad \text{\cite{CMS:2025fmx}}~,\\
    \Br(D^+ \to \pi^+ \mu \mu) <& 8.0 \times 10^{-8},\, ~ 95\%{\rm CL} \quad \text{\cite{LHCb:2020car}}~.\\
\end{split}\end{equation}
The predicted SM rates for these decays are negligible given the current and the future expected sensitivity, as discussed in Ref.~\cite{Gisbert:2020vjx}.
The same reference also reports the EFT dependence of both decays.
The expected sensitivity by the end of HL-LHC for the leptonic and semi-leptonic mode is $1.8 \times 10^{-10}$ and $3.7 \times 10^{-9}$, respectively, at 90\% CL \cite{Cerri:2018ypt}.

\subsection*{$\boldsymbol{s \to d \mu \mu}$}

The low-energy observable most sensitive to this flavor transition is the leptonic decay $K_{L} \to \mu \mu$. However, the SM prediction for this mode is hindered by long-distance non-perturbative contribution from the $K_L \to \gamma^*\gamma^*$ form factor (with off-shell photons), which is of comparable size to the clean short-distance contribution. By leveraging other Kaon decays, Ref.~\cite{Isidori:2003ts} put an upper limit on the short-distance term:
\be
    \Br(K_L \to \mu \mu)_{\rm SD} < 2.5 \times 10^{-9}~.
\ee
The EFT dependence is analogous to other leptonic decays and is reported explicitly in Ref.~\cite{Buchalla:1993wq}.

\subsection*{\boldsymbol{$B \to K \nu \nu}$ and $\boldsymbol{K \to \pi \nu \nu}$}

At tree-level, these decays are sensitive only to the four-fermion coefficients $C_{lq}^{(1)} - C_{lq}^{(3)}$ and $C_{ld}$.
For $B \to K \nu \nu$, the most precise measurement to date is the 2023 one from Belle-II \cite{Belle-II:2023esi}. The combination with previous measurements of the same mode is
\be
    \Br(B^+ \to K^+ \nu \nu) = (1.3 \pm 0.4) \times 10^{-5}~,
\ee
to be compared with the SM prediction $\Br(B^+ \to K^+ \nu \nu)_{\rm SM} = (4.44 \pm 0.30) \times 10^{-6}$ \cite{Becirevic:2023aov}.
Combining the two, we define the ratio
\be
    R^{\nu}_{K^+} = \frac{\Br(B^+ \to K^+ \nu \nu)}{\Br(B^+ \to K^+ \nu \nu)_{\rm SM}} = 2.93 \pm 0.90~.
\ee
With 50~ab$^{-1}$ of integrated luminosity, the Belle-II experiment expects to bring down the precision on this rare decay mode to 11\% of the SM rate \cite{Belle-II:2018jsg}.

The latest measurement of the very rare decay $K^+ \to \pi^+ \nu \nu$ from NA62 reads \cite{NA62:2024pjp}
\be
    \Br(K^+ \to \pi^+ \nu \nu) = (13.0 \pm 3.1) \times 10^{-11}~.
\ee
We take the SM prediction and EFT dependence from Refs.~\cite{Buras:2015qea,Buras:2015yca}, $\Br(K^+ \to \pi^+ \nu \nu)_{\rm SM} \approx 8.64 \times 10^{-11}$, and define the ratio
\be
    R_{K \to \pi \nu \nu} = \frac{\Br(K^+ \to \pi^+ \nu \nu)}{\Br(K^+ \to \pi^+ \nu \nu)_{\rm SM}} = 1.57 \pm 0.36~.
\ee
NA62 is expected to reach a precision of 15\% on the SM branching ratio by the end of its final run \cite{HIKE:2023ext}.

\subsection*{$\boldsymbol{t \to c \mu \mu}$ and $\boldsymbol{t \to u \mu \mu}$}

An ATLAS search for FCNC top decays $t \to Z q$ \cite{ATLAS:2018zsq} has been recast in Ref.~\cite{Chala:2018agk} to set limits on semi-leptonic SMEFT operators.
The resulting bounds the EFT scale $\Lambda$ are well below 1\,TeV, and are expected to reach only $\mathcal{O}(1 \TeV)$ with the full HL-LHC dataset.

\subsection*{$\boldsymbol{\tau \to 3 \mu}$}

A recent update from Belle-II \cite{Belle-II:2024sce} puts the following upper limit on this LFV decay:
\be
    \Br(\tau^- \to \mu^- \mu^+ \mu^-) < 2.3 \times 10^{-8}, \, 95\%{\rm CL}~.
\ee
The EFT dependence can be found, e.g., in Refs.~\cite{Kuno:1999jp,Crivellin:2013hpa}.
Belle-II is expected to reach a sensitivity of $3.3 \times 10^{-10}$ with 50~ab$^{-1}$ of integrated luminosity \cite{Belle-II:2018jsg}.

\subsection*{$\boldsymbol{\tau \to e \mu \mu}$}

The most recent constraint on this decay mode is from a 2010 Belle analysis \cite{Hayasaka:2010np}:
\be
    \Br(\tau^- \to e^- \mu^+ \mu^-) < 3.2 \times 10^{-8}, \, 95\%{\rm CL}~.
\ee
The EFT dependence can be found in Ref.~\cite{Crivellin:2013hpa}.
The final expected sensitivity from Belle-II is $4.0 \times 10^{-10}$ \cite{Belle-II:2018jsg,Cerri:2018ypt}.

\subsection*{LFU tests in $\boldsymbol{\tau}$ decays}
\label{sec:tauLFU}

At low energy, some of the strongest tests of LFU are obtained by taking ratios of leptonic tau and muon decays \cite{Stugu:1998jv,Pich:2013lsa}, currently tested at the per mille level \cite{HeavyFlavorAveragingGroupHFLAV:2024ctg}.
At the tree level, these ratios depend on the following combinations of SMEFT coefficients \cite{Allwicher:2021ndi}:
\begin{equation}\begin{split}
    \left|\frac{g_\mu}{g_e}\right|^2 &\equiv \frac{\Gamma(\tau \to \mu \nu\nu)}{\Gamma(\tau \to e \nu\nu)}  \left[\frac{\Gamma_{\rm SM}(\tau \to \mu \nu\nu)}{\Gamma_{\rm SM}(\tau \to e \nu\nu)} \right]^{-1} = \\  
        & = \frac{\left| \sqrt{2}G_F + C_{Hl}^{(3)[22]} + C_{Hl}^{(3)[33]} - 2 C_{ll}^{[2332]} \right|^2 + \left|C_{le}^{[2332]}\right|^2}
        { \left| \sqrt{2}G_F + C_{Hl}^{(3)[11]} + C_{Hl}^{(3)[33]} - 2 C_{ll}^{[1331]} \right|^2 + \left|C_{le}^{[1331]}\right|^2} = 1.0002 \pm 0.0022~, \\
    \left|\frac{g_\tau}{g_\mu}\right|^2 &\equiv \frac{\Gamma(\tau \to e \nu\nu)}{\Gamma(\mu \to e \nu\nu)} 
    \left[\frac{\Gamma_{\rm SM}(\tau \to e \nu\nu)}{\Gamma_{\rm SM}(\mu \to e \nu\nu)} \right]^{-1} = \\ 
        & = \frac{ \left| \sqrt{2}G_F + C_{Hl}^{(3)[11]} + C_{Hl}^{(3)[33]} - 2  C_{ll}^{[1331]} \right|^2 + \left|C_{le}^{[1331]}\right|^2}
        { \left| \sqrt{2}G_F + C_{Hl}^{(3)[11]} + C_{Hl}^{(3)[22]} - 2  C_{ll}^{[1221]} \right|^2 + \left|C_{le}^{[1221]}\right|^2}  = 1.0032 \pm 0.0028~, \\
    \left|\frac{g_\tau}{g_e}\right|^2 &\equiv \frac{\Gamma(\tau \to \mu \nu\nu)}{\Gamma(\mu \to e \nu\nu)} \left[\frac{\Gamma_{\rm SM}(\tau \to \mu \nu\nu)}{\Gamma_{\rm SM}(\mu \to e \nu\nu)}  \right]^{-1} = \\ 
        & =  
        \frac{\left| \sqrt{2}G_F + C_{Hl}^{(3)[22]} + C_{Hl}^{(3)[33]} - 2  C_{ll}^{[2332]} \right|^2 + \left|C_{le}^{[2332]}\right|^2}
        { \left| \sqrt{2}G_F + C_{Hl}^{(3)[11]} + C_{Hl}^{(3)[22]} - 2  C_{ll}^{[1221]} \right|^2 + \left| C_{le}^{[1221]}\right|^2}  = 1.0036 \pm 0.0028~,
\end{split}\end{equation}
where $G_F$ is the Fermi constant. The correlations between the three ratios are:\\
$\rho(\mu/e,\tau/\mu) = 0.40$, $\rho(\mu/e, \tau/e) = 0.67$, $\rho(\tau/\mu, \tau/e) = -0.41$ \cite{HeavyFlavorAveragingGroupHFLAV:2024ctg}.
The Belle-II experiment is expected to improve the precision in some $\tau$ LFU ratios by a factor $\sim 2$ \cite{Belle-II:2022cgf}, but no definitive numbers are reported for all of them.

\subsection*{$e^+ e^- \to \mu^+ \mu^-$ and $ \to $ and $\overset{\scriptscriptstyle(-)}{\nu}_{\!\!\mu} e^- \to \overset{\scriptscriptstyle(-)}{\nu}_{\!\!\mu} e^-$}
\label{sec:FP4lepton}

Limits on flavor-preserving four-lepton operators with two muons and two electrons can be derived from $\mu^+ \mu^-$ production at LEP2 \cite{Babich:2002jb,ALEPH:2013dgf} and muon (anti-)neutrino scattering on electrons, from CHARM \cite{CHARM:1988tlj}, CHARM-II \cite{CHARM-II:1994dzw}, and BNL-E734 \cite{Ahrens:1990fp}.
Ref.~\cite{Falkowski:2015krw} performed a global analysis in the SM EFT of these datasets, providing also individual constraints on the Wilson coefficients we are interested in, specifically $C_{ll}^{[1122]}$, $C_{le}^{[1122]}$, $C_{le}^{[2211]}$, and $C_{ee}^{[1122]}$. The EFT scale $\Lambda$ these measurements are sensitive to, at the 68\% CL, is approximately $3.5 \,\TeV$, as reported with gray bars in Figure~\ref{fig:4lepton_LFU}.

\newpage

\section{Tables}
\label{app:tab}

\renewcommand{\thetable}{\thesection.\arabic{table}}
\renewcommand{\theHtable}{\thesection.\arabic{table}}
\setcounter{table}{0}

\begin{table}[h]
\begin{center}
\begin{tabular}{|c||c|c||c|c||c|c|}
\hline
$\sigma_{\rm Bkg}$ [ab] & \multicolumn{2}{c|}{$\mu^- e^+$}  & \multicolumn{2}{c|}{$\mu^- \tau^+$}  & \multicolumn{2}{c|}{$e^- \tau^+$} \\\hline\hline
Bin $\cos\theta_{\ell^-}$ & 3\,TeV & 10\,TeV & 3\,TeV & 10\,TeV & 3\,TeV & 10\,TeV \\ \hline
$[-0.940,-0.752]$ & 0.070 & 0.012 & 0.92 & 0.085 & 0.77 & 0.077 \\
$[-0.752,-0.564]$ & 0.22 & 0.025 & 1.3 & 0.12 & 1.3 & 0.11 \\
$[-0.564,-0.376]$ & 0.51 & 0.048 & 1.5 & 0.13 & 1.4 & 0.13 \\
$[-0.376,-0.188]$ & 1.1 & 0.12 & 2.1 & 0.19 & 2.0 & 0.19 \\
$[-0.188,0]$ & 2.4 & 0.26 & 3.2 & 0.32 & 2.9 & 0.28 \\
$[0,0.188]$ & 5.2 & 0.51 & 5.1 & 0.51 & 4.6 & 0.42 \\
$[0.188,0.376]$ & 10 & 1.0 & 9.0 & 0.92 & 7.6 & 0.70 \\
$[0.376,0.564]$ & 21 & 2.1 & 17 & 1.8 & 13 & 1.2 \\
$[0.564,0.752]$ & 48 & 5.1 & 37 & 4.3 & 26 & 2.4 \\
$[0.752,0.940]$ & 170 & 20 & 130 & 20 & 59 & 5.7 \\\hline
\end{tabular}
\end{center}
 \caption{Cross sections, in ab, of the reducible backgrounds to LFV processes for each $\cos \theta$ bin, as derived from the simulations described in the main text.}
 \label{tab:LFVbkg}
\end{table}

\begin{table}[h]
\begin{center}
\begin{tabular}{|c|c|c|c|}
\hline
Coefficient & $\Lambda_{\rm MuC_{10}}$ [TeV] & $\delta(B_s \to \mu \mu)$ & $\delta(B \to K \nu \nu)$ \\\hline
$C_{lq}^{(1)[2232]}$ & 121 & 1.9\% & 0.42\% \\
$C_{lq}^{(3)[2232]}$ & 141 & 1.4\% & 0.31\% \\
$C_{qe}^{[3222]}$ & 127 & 1.7\% & - \\
$C_{ed}^{[2232]}$ & 124 & 1.8\% & - \\
$C_{ld}^{[2232]}$ & 118 & 2.0\% & 1.4\% \\
$C_{ledq}^{[2223]}$ & 109 & 105\% & - \\
$C_{ledq}^{[2232]}$ & 109 & 105\% & - \\\hline
\hline
\multicolumn{1}{|c|}{Coefficient} & 
\multicolumn{1}{|c|}{\(\Lambda_{\rm MuC_{10}}\) [TeV]} & 
\multicolumn{2}{c|}{\(\delta(B_d \to \mu \mu)\)} \\\hline
$C_{lq}^{(1)[2231]}$ & 126 & \multicolumn{2}{c|}{8.0\%} \\
$C_{lq}^{(3)[2231]}$ & 139 & \multicolumn{2}{c|}{6.6\%} \\
$C_{qe}^{[3122]}$ & 133 & \multicolumn{2}{c|}{7.2\%} \\
$C_{ed}^{[2231]}$ & 124 & \multicolumn{2}{c|}{8.3\%} \\
$C_{ld}^{[2231]}$ & 118 & \multicolumn{2}{c|}{9.3\%} \\
$C_{ledq}^{[2213]}$ & 109 & \multicolumn{2}{c|}{990\%} \\
$C_{ledq}^{[2231]}$ & 109 & \multicolumn{2}{c|}{990\%} \\\hline
\end{tabular}
\end{center}
\caption{Sensitivity at 68\% CL of a 10\,TeV MuC for flavor-changing operators involving the bottom quark, from high-energy measurements. Also shown is the relative effect these values of the coefficients on the branching ratio of the rare $B$ decays.}
\label{tab:MuC_rareBdecays}
\end{table}

\begin{table}[h]
\begin{center}
\begin{tabular}{|c|c|c|c|c|}
\hline
Coefficient & $\Lambda_{\rm MuC_{10}}$ [TeV]  & $\text{Br}(\tau\to 3 \mu)$ \\\hline
$C_{ll}^{[2223]}$ & $ 187$ & $2.7 \times 10^{-13}$ \\
$C_{le}^{[2223]}$ & $ 166$ & $2.2 \times 10^{-13}$ \\
$C_{le}^{[2322]}$ & $ 166$ & $2.2 \times 10^{-13}$ \\
$C_{ee}^{[2223]}$ & $ 210$ & $1.7 \times 10^{-13}$ \\ \hline
\end{tabular}
\end{center}
 \caption{Sensitivity at 68\% CL of a 10\,TeV MuC for SMEFT coefficients contributing to LFV in $\tau \to 3\mu$ decay. We also show the effect those coefficients would have on the $\tau \to 3 \mu$ branching ratio.}
 \label{tab:MuC_tauLFV}
\end{table}

\begin{table}[h]
\begin{center}
\begin{tabular}{|c|c|c|c|c|}
\hline
Coefficient & $\Lambda_{\rm MuC_{10}}$ [TeV]  & $\delta(|g_\mu/g_e|^2)$ & $\delta(|g_\tau/g_\mu|^2)$  & $\delta(|g_\tau/g_e|^2)$   \\\hline
$C_{ll}^{[1221]}$ & 260 & 0 & $8.9 \times 10^{-7}$ & $8.9 \times 10^{-7}$ \\
$C_{ll}^{[2332]}$ & 205 & $1.4 \times 10^{-6}$ & 0 & $1.4 \times 10^{-6}$ \\
$C_{le}^{[1221]}$ & 88 & 0 & $1.5 \times 10^{-11}$ & $1.5 \times 10^{-11}$ \\
$C_{le}^{[2332]}$ & 75 & $2.9 \times 10^{-11}$ & 0 & $2.9 \times 10^{-11}$ \\\hline
$C_{Hl}^{(3)[22]}$ & 183 & $3.6 \times 10^{-6}$ & $3.6 \times 10^{-6}$  & $1.1 \times 10^{-6}$ \\\hline
\end{tabular}
\end{center}
 \caption{Sensitivity at 68\% CL of a 10\,TeV MuC for EFT operator coefficients contributing to both LFUV tests from $\tau$ decays and $\ell^+ \ell^-$ production at a MuC. We also show the corresponding relative effects these values of coefficients would have on the $\tau$ LFU tests.}
 \label{tab:MuC_tauLFU}
\end{table}

\bibliographystyle{JHEP}
\bibliography{biblio}

\end{document}